\documentclass[%
 reprint,
 amsmath,amssymb,
 aps,
]{revtex4-2}

\usepackage{xcolor}

\usepackage{graphicx}
\usepackage{dcolumn}
\usepackage{bm}
\usepackage[normalem]{ulem}

\usepackage[utf8]{inputenc}
\usepackage[T1]{fontenc}
\usepackage{mathptmx}
\usepackage{tikz}
\usepackage{xcolor}
\usepackage{siunitx}
\usepackage{enumerate}
\usetikzlibrary{decorations.pathmorphing}

\newcommand{\bx}{\mathbf{x}}
\newcommand{\bk}{\mathbf{k}}
\newcommand{\bp}{\mathbf{p}}
\newcommand{\br}{\mathbf{r}}

\newcommand{\bR}{\mathbf{R}}

\def\ee{{\rm e}}
\def\ii{{\rm i}}

\definecolor{ggreen}{RGB}{50,172,58}
\definecolor{bblue}{RGB}{35,139,193}

\begin{document}

\preprint{APS/123-QED}

\title{Inelastic Mach-Zehnder Interferometry with Free Electrons}

\author{Cameron W. Johnson$^1$}%
 \email{cwj@uoregon.edu}
\author{Amy E. Turner$^1$}%
\author{F. Javier Garc\'ia de Abajo$^{2,3}$}%
\author{Benjamin J. McMorran$^1$}%
\affiliation{$^1$Department of Physics, University of Oregon, Eugene, Oregon 97403, USA}
\affiliation{$^2$ICFO-Institut de Ciencies Fotoniques, Mediterranean Technology Park, 08860 Castelldefels (Barcelona), Spain}
\affiliation{$^3$ICREA-Instituci\'o Catalana de Recerca i Estudis Avan\c{c}ats, Passeig Llu\'{\i}s Companys 23, 08010 Barcelona, Spain}

\date{\today}

\begin{abstract}
We use a novel scanning electron Mach-Zehnder interferometer constructed in a conventional transmission electron microscope to perform inelastic interferometric imaging with free electrons. An electron wave function is prepared in two paths that pass on opposite sides of a gold nanoparticle, where plasmons are excited before the paths are recombined to produce electron interference. We show that the measured spectra are consistent with theoretical predictions, specifically that the interference signal formed by inelastically scattered electrons is pi out of phase with respect to that formed by elastically scattered electrons. This technique is sensitive to the phase of localized optical modes because the interference signal amounts to a substantial fraction of the transmitted electrons. We thus argue that inelastic interferometric imaging with our scanning electron Mach-Zehnder interferometer provides a new platform for controlling the transverse momentum of a free electron and studying coherent electron-matter interactions at the nanoscale.
\end{abstract}

\maketitle


\emph{Introduction.---} 
Free electrons in a transmission electron microscope (TEM) are ideal for probing individual, nanoscale plasmonic systems \cite{garcia_de_abajo_optical_2010} due to their ability to couple with electromagnetic fields and form high resolution images. These interactions can be measured with electron energy-loss spectroscopy (EELS), which is sensitive to the photonic local density of states (i.e., the field intensity of the probed optical modes) but insensitive to the phase of the excited fields \cite{garcia_de_abajo_optical_2010}. In general, polaritonic excitations (plasmons) and their radiated fields are phase coherent, and passing free electrons can scatter and preserve the plasmon's phase information. Thus, scattered free electrons are viable quantum probes for manipulating and measuring nanoplasmonic systems \cite{polman_electron-beam_2019}. Here we explore the relationship between the phase of both free electrons and the plasmons they generate.

Several experiments in TEMs have exploited this phase coherence associated with inelastic electron-matter interactions. In particular, inelastic holography is an interferometric technique using an electrostatic biprism to interfere different parts of an electron wave after interacting with the sample, specifically looking at the coherent interference of the inelastically scattered electrons. This approach has measured the coherence properties of bulk and surface plasmon excitations \cite{lichte_inelastic_2000,roder_inelastic_2011}, as well as the loss of coherence due to the electromagnetic interaction with thermally populated material excitations \cite{kerker_quantum_2020}. However, inelastic holoography requires a high degree of electron spatial coherence. The partial coherence of the electron source and the production of multiple final scattering states after electron-sample interactions complicate the analysis and interpretation of measured signals \cite{chang_optimising_2015,verbeeck_plasmon_2005,verbeeck_interpretation_2006}. An alternative method to holography is wavefront matching. The wavefront of a single electron can be shaped to match the spatial extent and phase of the plasmonic near field. Then, post-selecting the coherently scattered wavefront in the far field can be used to spectrally filter for specific plasmonic modes. A successful demonstration of wavefront selection consisted in filtering the dipolar localized plasmon resonance (LPR) from the quadrupolar LPR of a metallic nanorod \cite{guzzinati_probing_2017}. This method has also been proposed to measure the transfer of orbital angular momentum \cite{ugarte_controlling_2016}. However, post-selecting wavefronts is very inefficient and requires precise alignments in the mode matching and selection apertures that limit acquisition of high-quality images in real-time. Alternatively, plasmonic interactions can be resonantly enhanced by temporally and spatially matching pulses of probe electrons with plasmonic near fields excited by an external optical pump \cite{barwick_photon-induced_2009}. This resonance enhancement method has successfully excited a plasmon and detected the plasmonic near field using a matched longitudinal electron wave function \cite{rivera_lightmatter_2020}, but requires highly specialized TEMs and ultrafast optical systems. The robust electron Mach-Zehnder interferometer described in this letter can improve on many of these approaches due to its ability to scan spatially separated paths, tune the phase, and create discrete, co-propagating outputs. The novel inelastic interferometer opens doors to a diverse range of interferometric experiments that were not possible before.

Mach-Zehnder interferometers for electrons have existed for quite some time \cite{marton_electron_1952,marton_electron_1954,gronniger_three-grating_2006} and were recently implemented in a TEM using monolithic crystals \cite{tavabi_new_2017,agarwal_nanofabricated_2017}. More recently, improvements in the fabrication of electron apertures featuring efficient phase gratings \cite{johnson_improved_2020} have enabled the implementation of a flexible and tunable two-grating electron Mach-Zehnder interferometer (2GeMZI) in a conventional TEM \cite{johnson_scanning_aps_2021}. The 2GeMZI is constructed with an amplitude-dividing beamsplitter grating \cite{yasin_path-separated_2018} forming tightly focused probes to interrogate a sample. After interaction with the sample, the probes pass through a second grating, recombining the separate paths for co-propagation to a detector. With flexible control over the relative probe positions and phases, the 2GeMZI can conveniently match plasmon resonance modes. Co-propagation in the interferometer output allows for complete interference, mitigating the need for post-selection apertures, reducing the complexity of the analysis, and dramatically increasing the interference signal to order unity. Probing the scattering from a plasmonic near field using the 2GeMZI as an inelastic interferometer combines the concepts of transverse beam-shaping and inelastic holography, gaining all the capabilities of these individual techniques while mitigating their faults. All of this can be accomplished by simply modifying two apertures in a conventional TEM.

In this Letter, we use a two-probe 2GeMZI to image the interference of coherent superpositions of electrons inelastically scattered by the self-induced LPR excitations of an individual spherical gold nanoparticle (NP). The interferometer output is collected by an EELS system where the plasmon scattered electrons can be spectrally resolved from the zero loss peak (ZLP) (Fig.~\ref{fig:fig1}). The relative phase between the two-probes in the interferometer output can be arbitrarily tuned by shifting the incident image of the input grating (G1) with respect to the output grating (G2), or by passing the probes through a spatially varying electrostatic potential. We tune the interferometer phase to enhance and suppress the dipole LPR spectral peak in the interferometer's output, which we show is pi out of phase with the elastically scattered signal. The constructive and destructive interference of the inelastic signal as a function of the interferometer phase demonstrates that electrons in a coherent superposition of paths can interact inelastically with the sample and still retain coherence.

\begin{center}
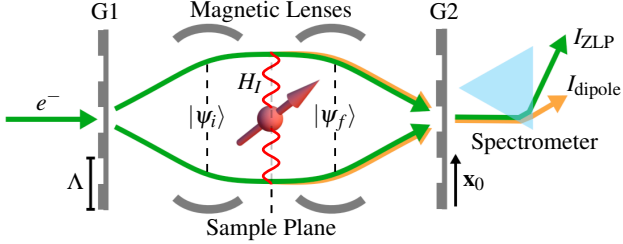
\begin{figure}[h]
\begin{tikzpicture}
    \node[inner sep=0pt] at (0,0)
        {\includegraphics[width=3.25in]{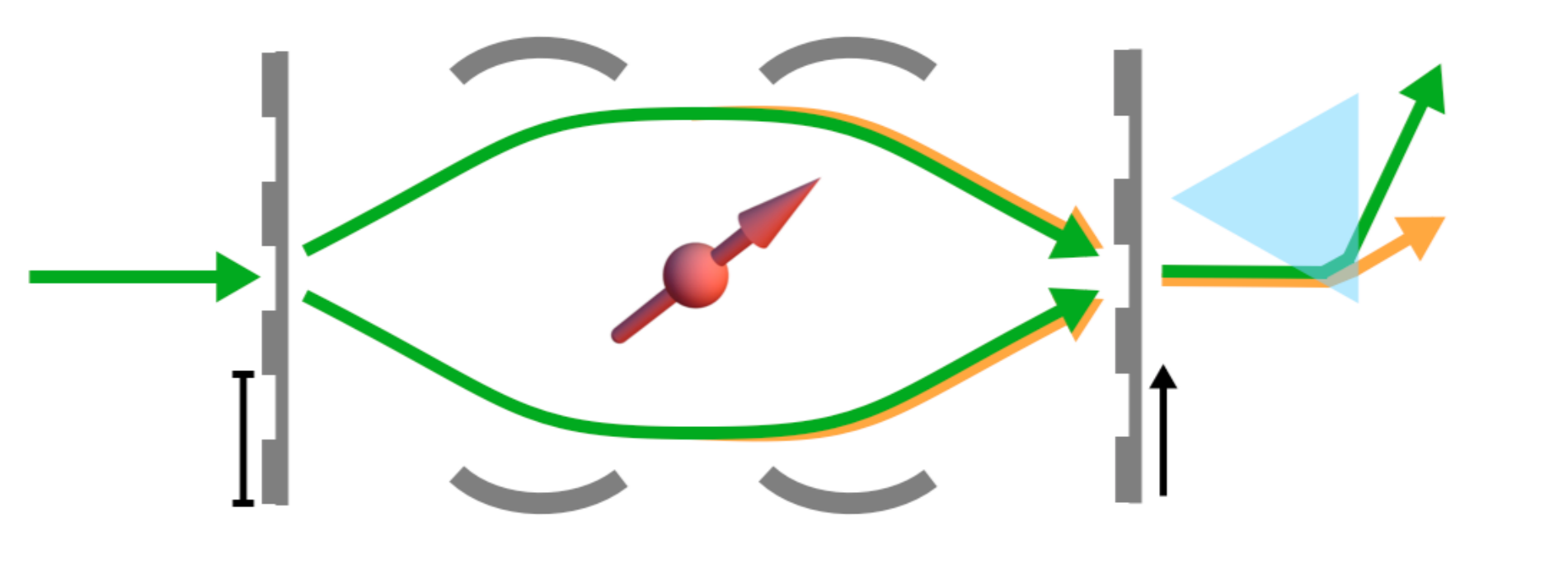}};
    \node[scale=1,inner sep=0pt] at (3.85,1.1)
        {$I_{\text{ZLP}}$};
    \node[scale=1,inner sep=0pt] at (3.85,0.45)
        {$I_{\text{dipole}}$};
    \draw[-,line width=1.0pt,red,decorate,decoration={snake,amplitude=3pt,pre length=0pt,post length=0pt}] (-0.45,-0.095) -- ++(0,-0.749);
    \draw[-,line width=1.0pt,red,decorate,decoration={snake,amplitude=3pt,pre length=0pt,post length=0pt}] (-0.45,0.17) -- ++(0,0.75);
    \node[scale=1,inner sep=0pt] at (-3.05,-0.8)
        {$\Lambda$};
    \node[scale=1,inner sep=0pt] at (-3.4,0.275)
        {$e^-$};
    \node[scale=1,inner sep=0pt] at (2.25,-0.825)
        {$\bx_0$};
    \node[scale=1,inner sep=0pt] at (-1.3,0.045)
        {$|\psi_i\rangle$};
    \draw[-,line width=0.7pt,dashed] (-1.3,0.225) -- ++(0,0.6);
    \draw[-,line width=0.7pt,dashed] (-1.3,-0.15) -- ++(0,-0.6);
    \node[scale=1,inner sep=0pt] at (-0.725,0.5)
        {$H_I$};
    \node[scale=1,inner sep=0pt] at (0.4,0.045)
        {$|\psi_f\rangle$};
    \draw[-,line width=0.7pt,dashed] (0.4,0.225) -- ++(0,0.6);
    \draw[-,line width=0.7pt,dashed] (0.4,-0.15) -- ++(0,-0.6);
    \draw[-,line width=0.7pt,dashed] (-0.445,-1.225) -- ++(0,0.4);
    \draw[-,opacity=0.25,line width=0.7pt,dashed] (-0.445,-1.225) -- ++(0,2.1);
    \node[scale=1,inner sep=0pt] at (-0.45,-1.45)
        {Sample Plane};
    \node[scale=1,inner sep=0pt] at (-2.65,1.46)
        {G1};
    \node[scale=1,inner sep=0pt] at (-0.45,1.45)
        {Magnetic Lenses};
    \node[scale=1,inner sep=0pt] at (1.85,1.46)
        {G2};
    \node[scale=1,inner sep=0pt] at (3.05,-0.3)
        {Spectrometer};
\end{tikzpicture}
\caption{\label{fig:fig1} Illustration of experiment. An electron is split by a grating (G1) with pitch $\Lambda=2\pi/|\bk_0|$, preparing it in the superposition state $|\psi_i\rangle$. The separated paths are focused and steered by magnetic lenses to interact with the dipolar plasmon of a gold nanoparticle (interaction Hamiltonian $H_I$) giving the final state $|\psi_f\rangle$. The paths are then recombined using a second grating (G2 with the same pitch $\Lambda$) that is continuously translatable by $\bx_0$. The position $\bx_0$ of G2 provides a way to tune the relative path phase in the interferometer output as $\Delta\phi_\text{int}=2\bk_0\cdot\bx_0$. The output of the interferometer is then dispersed in a spectrometer to resolve the elastic ZLP (green) and a dipole plasmon-scattered peak (orange). }
\end{figure}
\end{center}

\emph{Dipolar Interactions in the 2GeMZI.---} The 2GeMZI separates the electron wavefuntion into two parts, as depicted in Fig. \ref{fig:fig1}. The incident electron state $|\psi_i\rangle$, is prepared such that there are two-focal-spots at the sample plane that are physically created using the first grating, G1. The final state of the electron $|\psi_f\rangle$ is post-selected after interacting with the sample by means of a second grating, G2, and the entrance aperture of the electron energy analyzer. Near the sample plane, we can write the real-space representation of the incident electron wave function as $\psi_i(\br)=\langle\br|\psi_i\rangle=\left[\chi_i(\br-\br_1)+\ee^{\ii\phi_i}\chi_i(\br-\br_2)\right]/\sqrt{2}$, where $\chi_i$ is a normalized function describing each of the two nonoverlapping spots (centered at positions $\br_1$ and $\br_2$, respectively) and $\phi_i$ is a relative phase shift produced by stray potentials or any other differences between the two electron paths, such as the alignment of grating G1 relative to the optical axis (see below). Ignoring off-axis diffracted beams after grating G2, the state $|\psi_f\rangle$ takes the same functional form as the time reversal of $|\psi_i\rangle$ (i.e., it can be regarded as the mirror image of $|\psi_i\rangle$ through the sample plane, but with the electron traveling backwards). Consequently, we can write $\psi_f(\br)=\langle\br|\psi_f\rangle=\left[\chi_f(\br-\br_1)+\ee^{\ii\phi_f}\chi_f(\br-\br_2)\right]/\sqrt{2}$. Additionally, for a lateral displacement $\bx_0$ of grating G1 (or G2) along a direction across the grooves, we have that $\phi_i$ (or $\phi_f$) is modified by a term $2\bk_0\cdot\bx_0$, where $|\bk_0|=2\pi/\Lambda$ is the wavenumber of the gratings with pitch $\Lambda$. This terms arises directly from the phase difference between the two first-order diffraction orders introduced by a translation of the grating mask.

In the absence of a sample, the arguments above allow us to write the measured electron intensity output of the interferometer (i.e., the ZLP) as
\begin{align}
I_{\rm ZLP}&\propto|\langle\psi_f|\psi_i\rangle|^2\propto\cos^2(\Delta\phi/2), \label{eq:eq1} 
\end{align}
where the total relative path phase $\Delta\phi = \Delta\phi_\text{int}+\Delta\phi_\text{ext}$ receives contributions from the interferometer alignment $\Delta\phi_\text{int}=2\bk_0\cdot\bx_0$ and the noted path-related phases $\Delta\phi_\text{ext}=\phi_f-\phi_i$. In contrast, when a sample is inserted, the inelastic signal becomes $\propto\sum_e|\langle\psi_f|\langle e|H_I|g\rangle|\psi_i\rangle|^2$, where $H_I$ is the electron-sample interaction Hamiltonian, $|g\rangle$ represents the initial sample ground state, we sum incoherently over all final excited sample states $|e\rangle$, and each term in the sum involves a different final electron energy. However, the electron wave function can still be approximated by $\psi_f(\br)$ for each $|e\rangle$ (i.e., the energy loss does not significantly affect electron propagation, other than in the spectral separation performed at the analyzer \cite{krehl_spectral_2018}). For a dipolar sample excitation of transition dipole $\bp$ placed at a position $\br_c=(\br_1+\br_2)/2$ (the sample is centered between the two electron spots) and oriented along the interspot direction $\br_2-\br_1$, we have $H_I(\br)\propto\bp\cdot(\br-\br_c)$. So, for a small spot size compared to $r_c$, the inelastic signal becomes
\begin{align}
I_{\rm dipole}\propto\sin^2(\Delta\phi/2). \label{eq:eq2} 
\end{align}
In general, for an excitation characterized by an angular momentum number $m$, we have $H_I(\br)\propto\ee^{\ii m\varphi}$, where $\varphi$ is the azimuthal angle relative to $\br_c$. Following the same procedure as above, the resulting inelastic signal is then $\propto\cos^2[(\Delta\phi+m\Delta\varphi)/2]$, where $\Delta\varphi=\varphi_2-\varphi_1$ is the relative azimuthal angle between the two electron positions with respect to $\br_c$. By directly applying this analysis to each multipole of a spherical particle, we find an inelastic plasmon 2GeMZI signal
\begin{equation}
    \begin{split}
        I_{\rm sphere}(\Delta\phi) \propto &\frac{e^2}{\hbar\omega c}\sum_{l=1}^\infty\sum_{m=-l}^l
        C_{lm}^E\mbox{Im}\{t_l^E(\omega)\}K_m^2\left(\frac{\omega R}{v\gamma}\right) \\
        &\times\cos^2[(\Delta\phi+m\Delta\varphi)/2] \label{eq:eq3}
    \end{split}
\end{equation}
if the two spots are each at a radial distance $R$ from the sphere's center (see in SI \cite{EPAPS2GeMZI}). In this expansion we only retain electric modes ($E$) that dominate the response of the NP, and in particular the $l=1$ terms are due to the dipole excitation, with $|m|=1$ corresponding to the sample-plane-oriented dipole and $m=0$ denoting the along-the-beam dipole. Furthermore, we define the interference part of the inelastic signal
\begin{equation}
    \begin{split}
        I_\text{interference} = I_{\rm sphere}(\Delta\phi_\text{int}=0)-I_{\rm sphere}(\Delta\phi_\text{int}=\pi) \label{eq:eq4}
    \end{split}
\end{equation}
(see below).

Experimentally, we probe spherical NPs with a $\approx60$ nm diameter and an 80 nm path separation in the 2GeMZI. With these experimental parameters, the $l=1$ dipole mode in Eq. (\ref{eq:eq3}) is dominant and the higher order terms can be safely neglected. Additionally, we have recently shown that multi-probe imaging with the 2GeMZI induces sample charging that can cause significant relative phase shifts between the interferometer paths \cite{johnson_scanning_aps_2021}. This external phase shift can be effectively modeled for the probes at transverse positions $\bR_1$ and $\bR_2$ passing through an electrostatic potential produced by the sample charging as $\Delta\phi_\text{ext}=\sigma[V_z(\bR_2)-V_z(\bR_1)]$, where $\sigma = e/\hbar v$ is the first-order interaction parameter of an electron with $v$ the electron velocity and $V_{z}(\bR) = \int dz\,V(\br)$ is the projected potential \cite{cowley_electron_1972} (see in SI \cite{EPAPS2GeMZI}).

\onecolumngrid
\begin{center}
\begin{figure}[h]
\begin{tikzpicture}
    \node[inner sep=0pt] at (0,0){\includegraphics[width=7in]{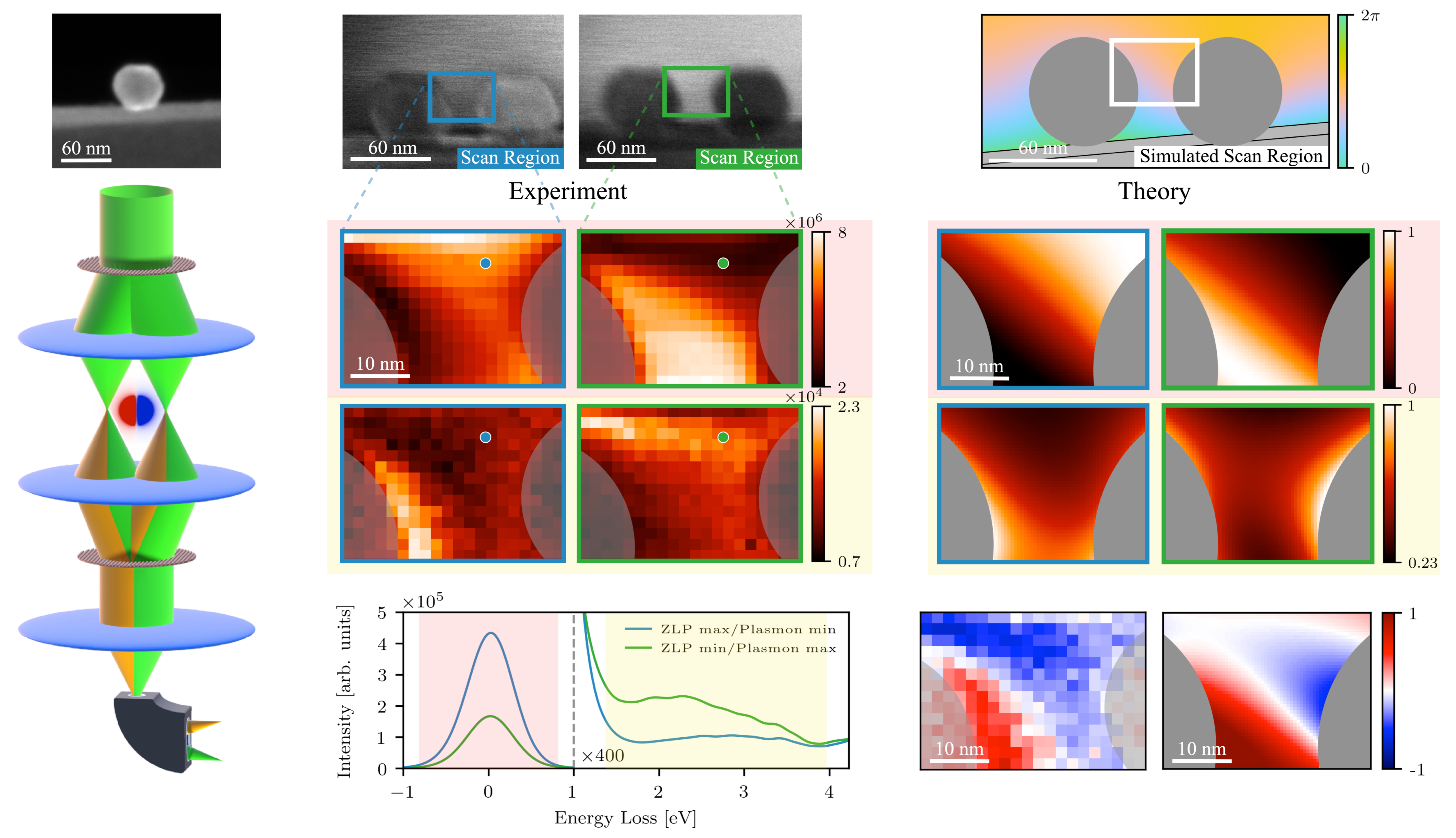}};
    \node[white,inner sep=0pt] at (-8.0,4.75){(a)};
    \node[black,inner sep=0pt] at (-8.5,2.55){(b)};
    \node[white,inner sep=0pt] at (-4.45,4.75){(c)};
    \node[white,inner sep=0pt] at (-1.55,4.75){(d)};
    \node[inner sep=0pt] at (3.35,4.75){(g)};
    \node[inner sep=0pt] at (-3.4,2.575){$\Delta\phi_\text{int}=\pi$};
    \node[inner sep=0pt] at (-0.6,2.575){$\Delta\phi_\text{int}=0$};
    \node[inner sep=0pt] at (3.9,2.575){$\Delta\phi_\text{int}=\pi$};
    \node[inner sep=0pt] at (6.5,2.575){$\Delta\phi_\text{int}=0$};
    \node[white,inner sep=0pt] at (-4.45,2.0){(e)};
    \node[white,inner sep=0pt] at (2.85,2.075){(h)};
    \node[inner sep=0pt] at (-3.7,-2.6){(f)};
    \node[white,inner sep=0pt] at (2.6,-2.6){(i)};
    \node[inner sep=0pt] at (-5.75,-3.5){Plasmon};
    \node[inner sep=0pt] at (-5.85,-4.3){ZLP};
    \node[inner sep=0pt] at (-8.3,1.9){G1};
    \node[inner sep=0pt] at (-8.3,1.0){Lens};
    \node[inner sep=0pt] at (-8.3,0.1){Sample};
    \node[inner sep=0pt] at (-8.3,-0.775){Lens};
    \node[inner sep=0pt] at (-8.3,-1.7){G2};
    \node[inner sep=0pt] at (-8.3,-2.565){Lens};
    \node[inner sep=0pt] at (-8.3,-3.6){EELS};
    \draw[->,line width=1.25pt] (-6.45,-1.675) -- ++(0.75,0);
    \node[inner sep=0pt] at (-6.05,-1.475){$\bx_0$};
    \node[scale=0.75,rotate=90,inner sep=0pt] at (1.4,1.3) {ZLP};
    \node[scale=0.75,rotate=90,inner sep=0pt] at (1.6,1.3) {Integrated};
    \node[scale=0.75,rotate=90,inner sep=0pt] at (1.4,1.3-2.15) {Plasmon};
    \node[scale=0.75,rotate=90,inner sep=0pt] at (1.6,1.3-2.15) {Integrated};
    \node[scale=0.75,rotate=90,inner sep=0pt] at (1.4+6.925,1.3) {ZLP};
    \node[scale=0.75,rotate=90,inner sep=0pt] at (1.6+6.925,1.3) {Integrated};
    \node[scale=0.75,rotate=90,inner sep=0pt] at (1.4+6.925,1.3-2.15) {Plasmon};
    \node[scale=0.75,rotate=90,inner sep=0pt] at (1.6+6.925,1.3-2.15) {Integrated};
    \node[rotate=90,inner sep=0pt] at (1.4+7.05,1.3-4.6) {$I_\text{interference}$};
\end{tikzpicture}
\caption{\label{fig:fig2} Demonstration of inelastic interferometry. (a) Dark field STEM image of a 60 nm gold nanoparticle (NP) isolated on the edge of a carbon substrate. (b) Sketch of the two-grating electron Mach Zehnder interferometer (2GeMZI), consisting of a STEM with two gratings used as beamsplitters. The first grating (G1) prepares electrons in a superposition of two separate paths, each of them interacting with the NP sample, with some probability of losing energy to a plasmon resonance (orange). (c,d) The electron paths are then recombined using the second grating (G2), which can be positioned for (c) destructive (blue borders) and (d) constructive (green borders) interference, conversely modifying the elastic and inelastic signals. Incidentally, two NPs are observed in the image because of the two-spot beam configuration, with the central frame selecting the interference region (i.e., each beam passing by one side of the NP). (e,f) For both alignment schemes, we integrate over the plasmon (yellow-shaded) and ZLP (red-shaded) regions of the energy loss spectra (f) at every scan location to create the spectral images shown in (e). The raw spectra in (f) correspond to the dotted positions in (e). (g,h) We simulate the experiment with an external potential $V_z(\bR)$ (g) and show that the calculated results (h) are qualitatively consistent with the experimental spectral images (e). (i) Interference term of the loss probability, obtained as the difference of the plasmon-integrated spectral images from the two different interferometer alignments in (e) and (h) [Eq. \ref{eq:eq4}].  }
\end{figure}
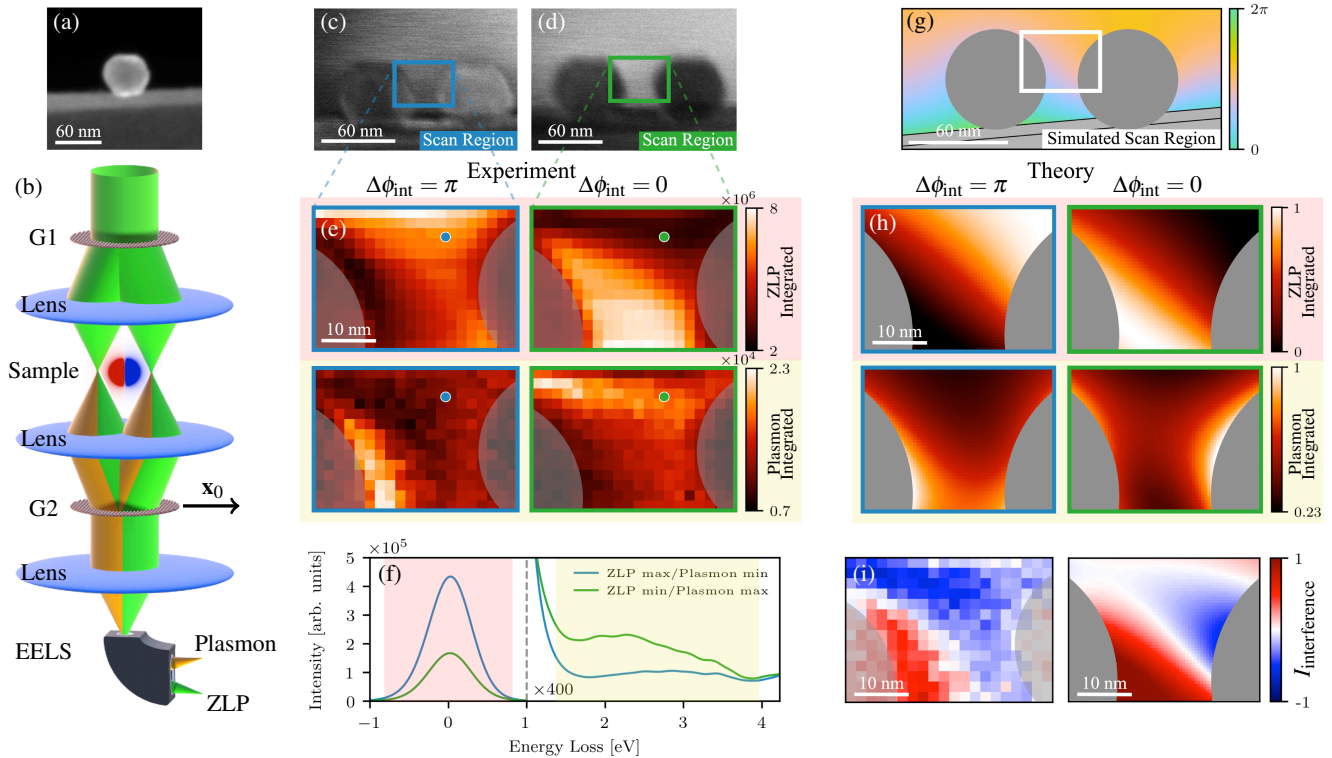
\end{center}
\twocolumngrid

\emph{Description of the Experiment.---} The 2GeMZI was constructed in an image corrected 80-300 keV FEI Titan TEM by placing the input grating G1 in the condenser 2 aperture holder above the specimen plane and the output grating G2 in the selected area aperture holder below the specimen plane, detailed description in \cite{johnson_scanning_aps_2021}. The TEM was operated at 80 keV in $\mu$probe mode, such that the STEM probe convergence angle was tunable from 1 to \SI{10}{\milli\radian}. Both G1 and G2 consisted of a 6$\times$6 array of \SI{30}{\micro\meter} diameter, 300 nm pitch binary diffraction gratings milled into a 30 nm thick free-standing Si$_3$N$_4$ membrane. These gratings were optimized to maximize intensity in the $\pm1$ diffraction orders while yielding a minimum in the zeroth diffraction order. Approximately 30\% of the total transmitted intensity was placed in each of the $\pm1$ orders, and no more than 6\% in any other diffraction order. The condenser 3 aperture \cite{johnson_scanning_aps_2021} was used to define the beam with a 3 mrad convergence angle, selecting a single grating from the array for G1. The Lorentz lens in the image corrector was then used to project a focused image of G1 onto G2. In addition, the post-G2 projection lenses were used to project a real-space image of the interferometer output into the entrance aperture of the EELS system. The natural energy spread of the emission source convolved with the point spread function (PSF) of the optical system of the spectrometer gave a measure of the full width at half maximum of the ZLP of 0.8 eV. In this configuration, the scan and descan coils were used to raster the probes in the specimen plane while keeping the image of G1 on G2 and the interferometer output on the EELS entrance aperture both stationary for up to a 200$\times$200 nm$^2$ scan region. The probe widths were 5 nm with a probe separation of 80 nm between the $\pm$1 diffraction orders.

Although the LPR intensity for a gold NP is more limited by spectral overlap with interband transitions than in other noble metals such as silver \cite{amendola_surface_2017}, gold was chosen for its resistance to form oxides, offering long stability, availability, and ease of sample preparation. A commercial 60 nm diameter monodisperse gold NP solution was dropcast on a lacey carbon grid, allowed to air dry, and then placed in the specimen plane of the TEM. A single NP was isolated on an edge of the carbon such that the two electron paths could pass on either side of the NP through vacuum [Fig. \ref{fig:fig2}(a,b)]. The two-probe scan regions were selected such that the probes were on either side of the NP with $\Delta\varphi\approx\pi$ [Fig. \ref{fig:fig2}(c,d)]. Then, spectral images were recorded for both destructive and constructive interferometer outputs [Fig. \ref{fig:fig2}(e)]. The imaging procedure was as follows: A larger-area bright-field scan was recorded to select a smaller region and collect a spectral image. The spectral image scan region was scanned while the interferometer output was collected by a 2 mm EELS entrance aperture with the spectrometer magnetic prism set to have an energy dispersion of 0.03 eV/channel on the CCD collecting the spectrum. The non-dispersing direction on the CCD was binned to 1 pixel and for each scan location 20 spectra were summed with each spectrum integrated for 0.01 seconds, giving a total integration time of 0.5 seconds per scan location. Once the spectral images were recorded, they were post-processed in two steps. First, each spectrum was smoothed with a 4-pixel standard deviation Gaussian convolution to remove high frequency noise. Second, 4 iterations of the 1D Richardson-Lucy deconvolution algorithm were run using the HyperSpy Python library on each spectrum. A vacuum-collected spectrum that was identically smoothed was used as the deconvolution kernel to partially remove the PSF of the spectrometer. This significantly narrowed the tails of the ZLP, effectively removing the background in the 1-3 eV region where the plasmon signals of interest were observed \cite{nelayah_mapping_2007}. Finally, the post-processed spectra were integrated over an energy range (-1,1) eV for the ZLP and (1.5,4) eV for the plasmon peaks, as shown in Fig. \ref{fig:fig2}(f). 

\emph{Results and Discussion.---} Figure \ref{fig:fig2}(a) shows that the carbon support has a much larger surface area than the NP and there is a small angular offset between the carbon edge and the horizontal diffraction direction of the scanning probes. Consequently, we model the contribution of the carbon edge to be 30 times stronger than that from the NP in the approximate external potential, and further account for a 5$^\circ$ angular misalignment between the probes and the carbon edge [Fig. \ref{fig:fig2}(g)]. We use this simulated potential to generate spectral images via Eq. (\ref{eq:eq3}) that are in excellent qualitative agreement with the experimental results [Fig. \ref{fig:fig2}(h)]. By taking the difference of the plasmon-integrated spectrum images with the destructive and constructive interferometer alignments, we can find the spatially resolved interference term in the energy-loss spectrum [Eq. (\ref{eq:eq4})] which shows the same structure as the simulated result [Fig. \ref{fig:fig2}(i)].

\begin{center}
\begin{figure}[h]
\begin{tikzpicture}
    \node[inner sep=0pt] at (0,0)
        {\includegraphics[width=3in]{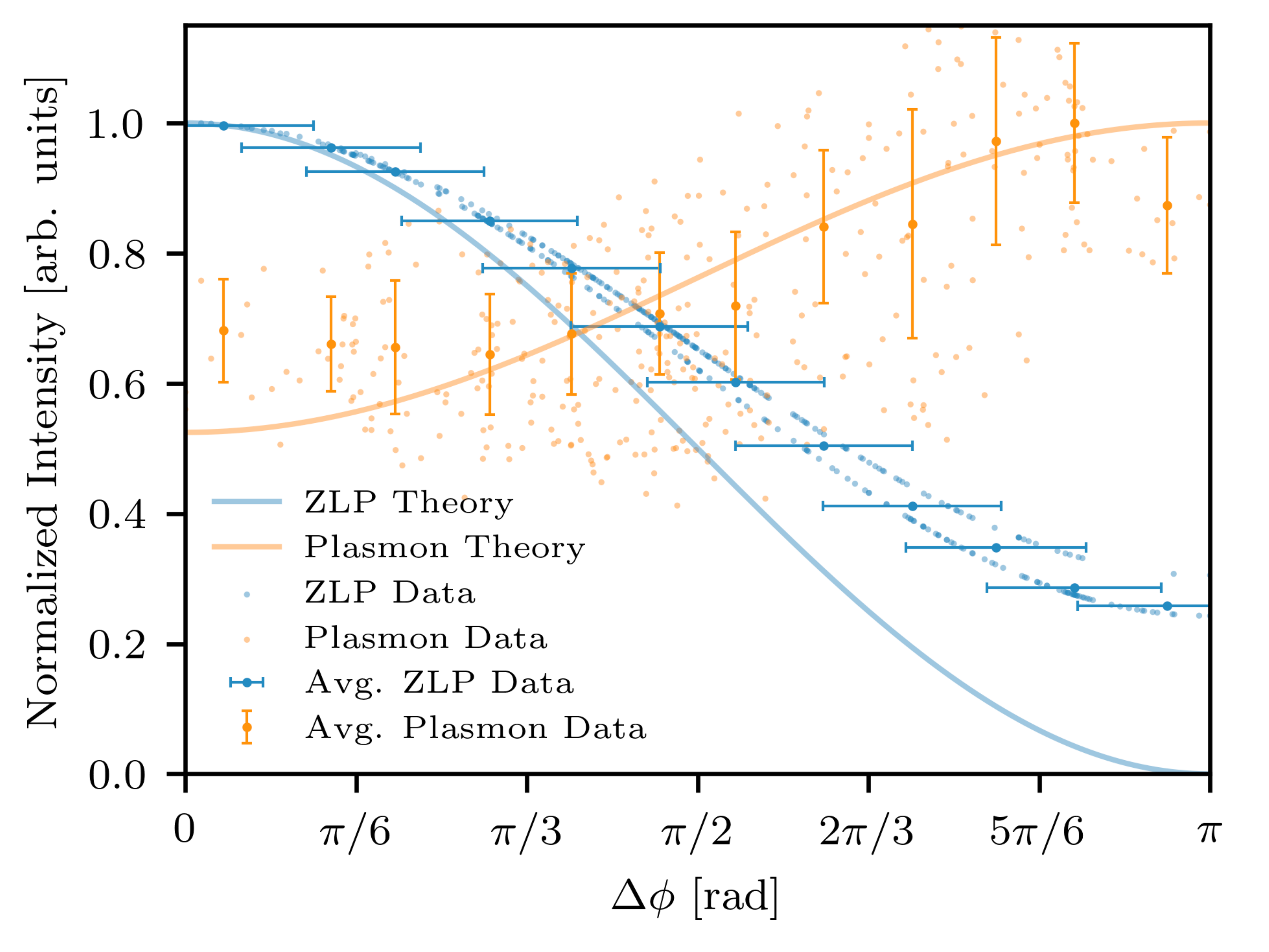}};
\end{tikzpicture}
\caption{\label{fig:fig3} Measured and modelled ZLP and plasmon intensities as a function of total relative probe phase. }
\end{figure}
\end{center}

To show the converse interference relation between the ZLP and plasmon signals more quantitatively, we assign a relative probe phase to each pixel with probes passing through vacuum in Fig. \ref{fig:fig2}(e) by normalizing the ZLP intensity, and then, we find the relative phase by inversion of Eq. (\ref{eq:eq1}). The normalized integrated ZLP and plasmon intensities of each pixel are plotted as a function of the assigned relative probe phase in Fig. \ref{fig:fig3}. For clarity, we also show the mean values binned by every $\pi/12$ relative phase interval along with the theoretically predicted values for $\Delta\varphi=\pi$. To account for our lack of knowledge of whether or not the minimum and maximum ZLP intensities correspond to the exact actual $\Delta\phi=\pi$ and $\Delta\phi=0$ points, we give a $\pi/12$ systematic error to the standard deviation of the binned values added in quadrature. Since we definitionally assigned the relative phase to the integrated intensity of the ZLP peak, we do not give an error for this integrated intensity. The error given for the mean integrated plasmon intensities is the standard deviation for each binned region and the phase error is assumed to be the same as the mean ZLP data. Deviation of the measured ZLP intensity from the theoretical prediction is well understood by the small, but nonzero contributions of the higher-order probes from G1 causing a loss of fringe visibility. This does not have a substantial effect on the visibility of the plasmon interference because the higher-order probes are further away from the NP than the main $\pm1$ probes and the plasmon loss probability is exponentially suppressed for large impact parameters [i.e., $K_m(x)\sim e^{-x}\sqrt{\pi/2x}$ for large $x$]. We made additional measurements on a separate gold NP that has an angular misalignment in the opposite direction. This dataset shows the spatial interference due to a skewed potential that is consistent with the opposing angular alignment and converse dipole interference relation (see in SI \cite{EPAPS2GeMZI}). 

Similar conditional interference relations exist between the ZLP and the higher-order modes in the multipole expansion, dependent on the geometry and symmetry of the mode's spatial distribution and probe positions, but the multipole plasmon peaks for a spherical gold NP of this size are not spectrally resolvable. In particular, the peak widths are much larger than the peak spacings in the loss spectrum. However, multiple plasmon mode peaks can be resolved with strongly anisotropic NPs \cite{bosman_mapping_2007}, chains of spherical NPs \cite{barrow_mapping_2014}, or by using less lossy materials \cite{kiewidt_numerical_2013}. These relations could be explored with this apparatus, but may require constructing another 2GeMZI in a monochromated TEM with an energy resolution below 100 meV. Incorporation of a cathodoluminescence collection system with a 2GeMZI could provide information about the correlation between the phase-coherent superpositions of scattered electrons to the radiated light from the dipole plasmon \cite{yamamoto_photon_2001,kociak_cathodoluminescence_2017}. Alternatively, this 2GeMZI could serve as an indirect way to measure the transfer of orbital angular momentum \cite{asenjo-garcia_dichroism_2014,zanfrognini_orbital_2019,tavabi_experimental_2021}.  Finally, we note that inelastic free electron interference in the 2GeMZI is not exclusive to plasmon scattering and can be used to probe other types of polaritons, as well as different condensed-matter quasiparticle excitations \cite{rivera_lightmatter_2020}.

\emph{Conclusion.---} We have demonstrated phase-sensitive interference between coherent superpositions of inelastically scattered free electrons within a two-probe 2GeMZI from plasmonic excitations of a single gold NP. The excitation of a plasmon introduces a relative $\pi$ phase difference between the two interferometer paths of the inelastically scattered electrons, which is well-described by a dipolar interaction. Elastically scattered electrons do not receive this phase shift. Thus the interferometer output of the plasmon loss channel is the complement of the ZLP channel. Isolating this interferometer output provides a robust way to detect dipole-excitations, even theoretically at energies lower than the energy resolution of the microscope. 

Beyond applications in probing nanoplasmonic systems, the high throughput, flexibility, scanning capabilities, and ease of operation in a conventional scanning TEM of this technique provides an exciting platform for probing quantum mechanics at the nanoscale, and additionally allows us to gain control over the transverse momentum of the free electron wave function. Further development of the techniques presented in this Letter could lead to tests of quantum complementarity for free electrons \cite{scully_quantum_1991}, explorations of decoherence theory \cite{schattschneider_entanglement_2018}, as well as the manipulation of free electrons with unprecedented versatility \cite{polman_electron-beam_2019}.

We thank Joshua Razink for the TEM instrument support. This work was supported by the NSF Grant No. 1607733 and NSF Grant No. 2012191. CWJ was supported by the NSF GRFP Grant No. 1309047. FJGA acknowledges support from the European Research Council (789104eNANO) the Spanish MINECO (PID2020-112625GB-I00 and SEV2015- 0522). 

\bibliographystyle{apsrev4-2}
\bibliography{ms.bib}

\end{document}


\renewcommand{\thefigure}{S\arabic{figure}}
\renewcommand{\theequation}{S\arabic{equation}}
\renewcommand{\thetable}{S\arabic{table}}
\renewcommand{\thesection}{S\arabic{section}}
	
\title{Inelastic Mach-Zehnder Interferometry with Free Electrons \\ {\color{gray}
\small -- SUPPLEMENTARY INFORMATION -- }
}

\author{Cameron W. Johnson} 
\email{cwj@uoregon.edu}
\affiliation{Department of Physics, University of Oregon, Eugene, Oregon 97403, USA}
\author{Amy E. Turner} 
\affiliation{Department of Physics, University of Oregon, Eugene, Oregon 97403, USA}
\author{F.~Javier~Garc\'{\i}a~de~Abajo} 
\affiliation{ICFO-Institut de Ciencies Fotoniques, The Barcelona Institute of Science and Technology, 08860 Castelldefels (Barcelona), Spain} 
\affiliation{ICREA-Instituci\'o Catalana de Recerca i Estudis Avan\c{c}ats, Passeig Llu\'{\i}s Companys 23, 08010 Barcelona, Spain}
\author{Benjamin J. McMorran} 
\affiliation{Department of Physics, University of Oregon, Eugene, Oregon 97403, USA}

\date{\today}


\maketitle


\section{Relativistic energy-loss probability from a Single free-electron Probe}

\begin{center}
\begin{figure}[h]
\begin{tikzpicture}
    \node[inner sep=0pt] at (0,0)
        {\includegraphics[width=3in]{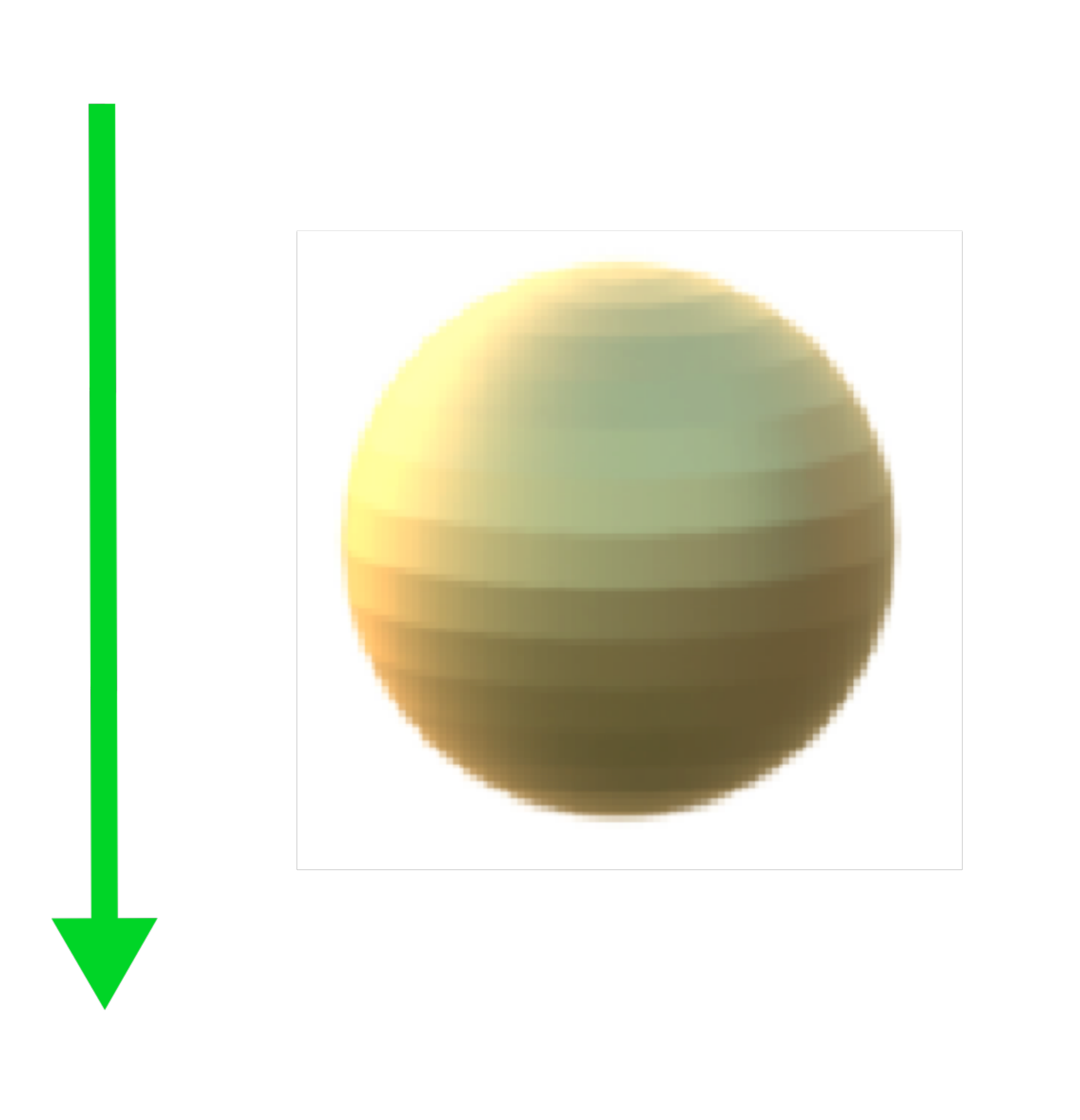}};
    \draw[|-|,line width=1]
    (-3.1,2.4) --
    (0.5,2.4);
    \node[inner sep=0pt] at (-1.5,2.7)
        {$R_0$};
    \draw[|-|,line width=1]
    (2,1.5) --
    (0.5,0.2);
    \node[inner sep=0pt] at (1.1,1.2)
        {$a$};
    \node[inner sep=0pt] at (-2.3,-2)
    {$\br_0(t)$};
    \draw[->,line width=2]
    (2.8,-2.25) --
    (2.8,-0.75);
    \draw[->,line width=2]
    (2.8,-2.25) --
    (3.35,-2.55);
    \draw[line width=0.25] (2.8,-2.25) ellipse (1.5 and 0.3);
    \draw[->,line width=2]
    (3.35,-2.55) --
    (4.8,-2.35);
    \node[inner sep=0pt] at (2.5,-2.3)
    {$\mathbf{\hat{R}}$};
    \node[inner sep=0pt] at (3.15,-1.4)
    {$\mathbf{\hat{z}}$};
    \node[inner sep=0pt] at (4.2,-2.75)
    {$\mathbf{\hat{\boldsymbol{\varphi}}}$};
\end{tikzpicture}
\caption{\label{fig:fig1} Geometry for a STEM probe with path described by $\br_0(t)$ at an impact parameter $R_0$ relative to a spherical metallic NP with radius $a$ in a cylindrical coordinate system.}
\end{figure}
\end{center}

\begin{center}
\begin{figure}[h]
\begin{tikzpicture}
    \node[inner sep=0pt] at (0,0)
        {\includegraphics[width=5in]{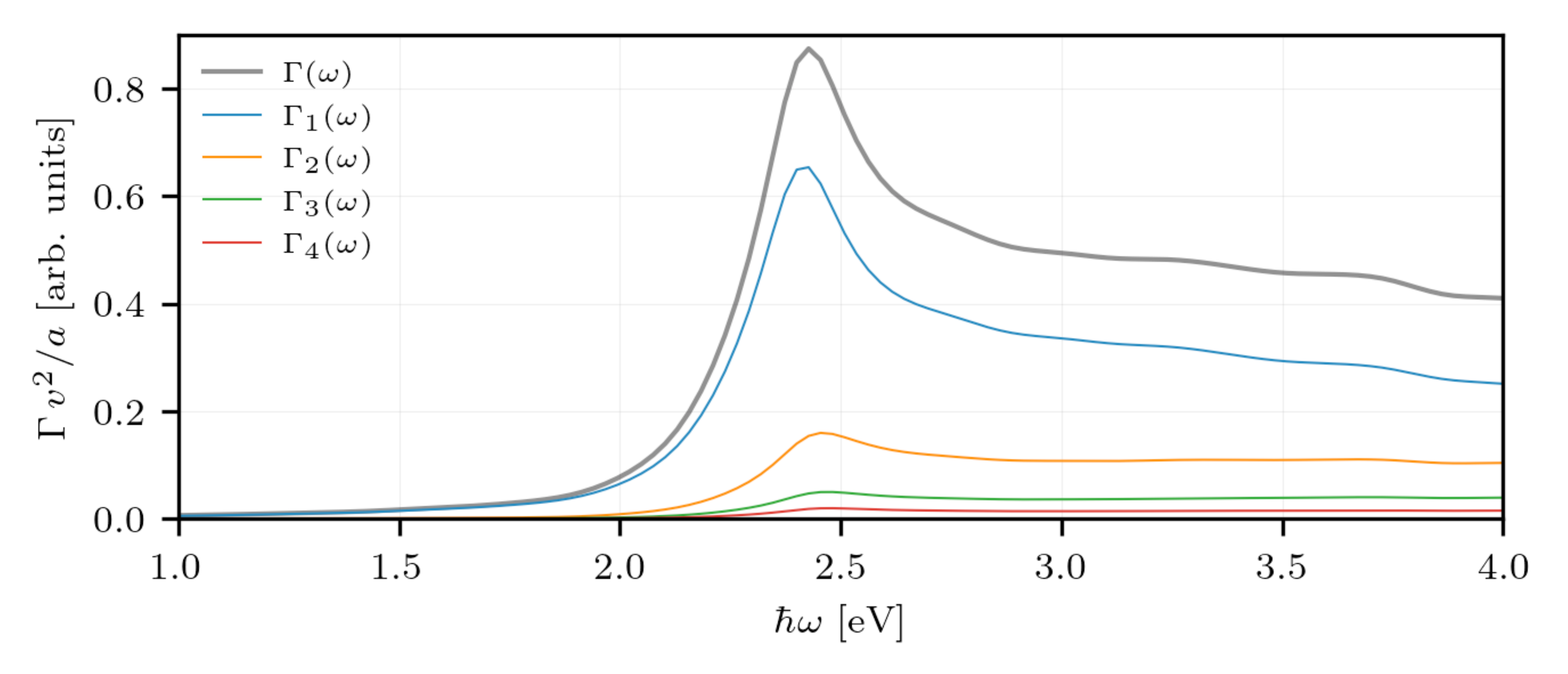}};
\end{tikzpicture}
\caption{\label{fig:fig2} Calculated single-probe energy-loss probability for a gold NP of $a=30$\,nm radius excited from a STEM probe at an impact parameter $R_0=1.2\,a$.}
\end{figure}
\end{center}

We consider a free-electron (FE) scanning transmission electron microscopy (STEM) probe passing near a spherical nanoparticle (NP) with radius $a$ at an impact parameter $R_0$, as illustrated in Fig.\ \ref{fig:fig1}. The probability of the FE losing an amount of energy $\hbar \omega$ by exciting a localized plasmon resonance while travelling at a velocity $v$ along a straight vertical trajectory $\br_0(t)=(R_0\cos(\varphi_0),R_0\sin(\varphi_0),z_0-vt)$ can be written as 
\begin{equation}
    \Gamma(\omega,\br_0,\br_1) = \frac{e}{\pi\hbar\omega}\int dt\, \mbox{Re}\left\{\ee^{-i\omega t}\bv\cdot\bE_{\br_1}^{\rm ind}(\br_0(t),\omega)\right\}, \label{eq:init_lossprob}
\end{equation}
where the electric field is induced by the dielectric response of the NP due to the external electric field of a FE following a the path $\br_1(t)$. Here, we have introduced different paths $\br_1(t)$ and $\br_0(t)$ for the exciting and the excited parts of the electron, respectively, which will be needed to describe the two-spot configuration in the 2GeMZI. For a conventional single-focus configuration, we need to take $\br_1(t)=\br_0(t)$. A fully relativistic analytical solution to Eq.\ (\ref{eq:init_lossprob}) can be found by expanding the induced electric field in terms of multipoles characterized by total and azimuthal angular momentum numbers $l$ and $m$, respectively. More precisely,
\begin{equation}
    \begin{split}
        \Gamma(\omega,\br_0,\br_1) = & \,\, \frac{e^2}{\hbar\omega c}\sum_{l=1}^\infty\sum_{m=-l}^lK_m\left(q_\gamma R_0\right)K_m\left(q_\gamma R_1\right) \mbox{Im}\Big\{\left[C^M_{lm}t^M_l(\omega)+C^E_{lm}t^E_l(\omega)\right] \\
        & \,\,\,\,\,\,\,\,\,\, \,\,\,\,\,\,\,\,\,\, \,\,\,\,\,\,\,\,\,\, \,\,\,\,\,\,\,\,\,\, \,\,\,\,\,\,\,\,\,\, \,\,\,\,\,\,\,\,\,\,  
        \,\,\,\,\,\,\,\,\,\, \,\,\,\,\,\,\,\,\,\, \,\,\,\,\,\,\,\,\,\, \,\,\,\,\,\,\,\,\,\, \,\,\,\,\,\,\,\,\,\, \,\,\,\,\,\,\,\,\,\,  
        \,\,\,\,\,\,\,\,\,\, \,\,\,\,\,\,\,\,\,\,
        \times \ee^{-im(\varphi_0-\varphi_1)-i\omega (z_0-z_1)/v} \Big\}, \\ \label{eq:2pEELS} 
    \end{split}
\end{equation}
where $K_m$ are modified Bessel functions of the second kind, $q_\gamma=\omega/v\gamma$ with $\gamma=1/\sqrt{1-v^2/c^2}$ the Lorentz factor and $c$ the speed of light in vacuum, $C^M_{lm}$ and $C^E_{lm}$ are magnetic and electric coupling coefficients that depend on $v$ but not on $\omega$, and $t_l^M$ and $t_l^E$ are the scattering matrix elements obtained from Mie scattering theory and explicitly given by
\begin{equation}
    \begin{split}
        t_l^M(\omega) = & \,\, \frac{-j_l(\rho_0)\rho_1j_l'(\rho_1)+\rho_0j_l'(\rho_0)j_l(\rho_1)}{h_l^{(+)}(\rho_0)\rho_1j_l'(\rho_1)-\rho_0[h_l^{(+)}(\rho_0)]'j_l(\rho_1)}, \\
        t_l^E(\omega) = & \,\, \frac{-j_l(\rho_0)[\rho_1j_l(\rho_1)]'+\epsilon[\rho_0j_l(\rho_0)]'j_l(\rho_1)}{h_l^{(+)}(\rho_0)[\rho_1j_l(\rho_1)]'-\epsilon[\rho_0h_l^{(+)}(\rho_0)]'j_l(\rho_1)}, \\
    \end{split}
\end{equation}
where $\rho_0=\omega a/c$, $\rho_1=\omega a\sqrt{\epsilon}/c$, $\epsilon$ is the dielectric function of the NP (taken from numerically tabulated measurements for gold \cite{johnson_optical_1972} in our calculations), $j_l$ and $h_l^{(+)}$ are spherical Bessel and Hankel functions, respectively, and the primes denote differentiation with respect to the argument. A full derivation of these expressions including closed forms for the coupling coefficients $C_{lm}^M$ and $C_{lm}^E$ can be found in Ref.\ \cite{garcia_de_abajo_relativistic_1999}. In particular, for the dominant electric dipole terms in our formalism, we have $C_{10}^E=(6/\pi)(c/v\gamma)^4$ and $C_{1,\pm1}^E=(3/\pi)(c^4/v^4\gamma^2)$. For a gold NP with 30\,nm radius and electrons of 80\,keV energy, $C^E_{lm}\mbox{Im}\{t^E_l\}\gg C^M_{lm}\mbox{Im}\{t^M_l\}$ in the 1-6\,eV spectral region, so we can safely neglect the magnetic contribution in the following analysis. In particular, the loss probability for the single-probe configuration reduces to
\begin{equation}
    \begin{split}
        \Gamma_{\{1\text{p}\}}(\omega,R_0) = & \,\, \frac{e^2}{\hbar\omega c}\sum_{l=1}^\infty\sum_{m=-l}^lK_m\left(q_\gamma R_0\right)^2 C^E_{lm}\mbox{Im}\{t^E_l(\omega)\} = \sum_{l=1}^\infty\Gamma^{\{1\text{p}\}}_l(\omega,R_0),
    \end{split}
\end{equation}
where the $\{1\text{p}\}$ subscript stands for one probe, and the $l=1,2,\dots$ terms correspond to the dipolar, quadrupolar, $\dots$ contributions to the loss probability. The different multipole components of the energy loss probability for this gold NP are plotted in Fig.\ \ref{fig:fig2}, showing that the LPR multipole peaks around 2.4 eV are not spectrally resolvable, and the dipole mode ($l=1$) produces the largest contribution to the spectrum. 

\FloatBarrier

In the two-focus configuration, we have $\br_0(t)\neq\br_1(t)$, but we take $z_0=z_1$, so Eq.\ (\ref{eq:2pEELS}) reduces to 
\begin{equation}
\label{GKCA}
    \begin{split}
        \Gamma(\omega,\br_0,\br_1) = & \,\, \frac{e^2}{\hbar\omega c}\sum_{l=1}^\infty\sum_{m=-l}^lK_m\left(q_\gamma R_0\right)K_m\left(q_\gamma R_1\right) C^E_{lm}\mbox{Im}\{t^E_l(\omega)\ee^{im(\varphi_0-\varphi_1))}\} \\
        \equiv &\,\, \mbox{Im}\{A(\omega,\bR_0,\bR_1)\}.
    \end{split}
\end{equation}
A complex loss function $A$ is implicitly defined in this expression, admitting the general form
\begin{equation}
\label{AAA}
    \begin{split}
        A(\omega,\bR_0,\bR_1)= -\frac{4e^2}{\hbar}\int dz\int dz'\ee^{i\omega(z-z')/v}\,G_{zz}(\omega,\bR_0,z,\bR_1,z')
    \end{split}
\end{equation}
in terms of the component $G_{zz}=\bz\cdot G\cdot\bz$ of the electromagnetic Green tensor \cite{garcia_de_abajo_optical_2010}.

\begin{center}
\begin{figure}[h]
\begin{tikzpicture}
    \node[inner sep=0pt] at (0,0)
        {\includegraphics[height=3.0in]{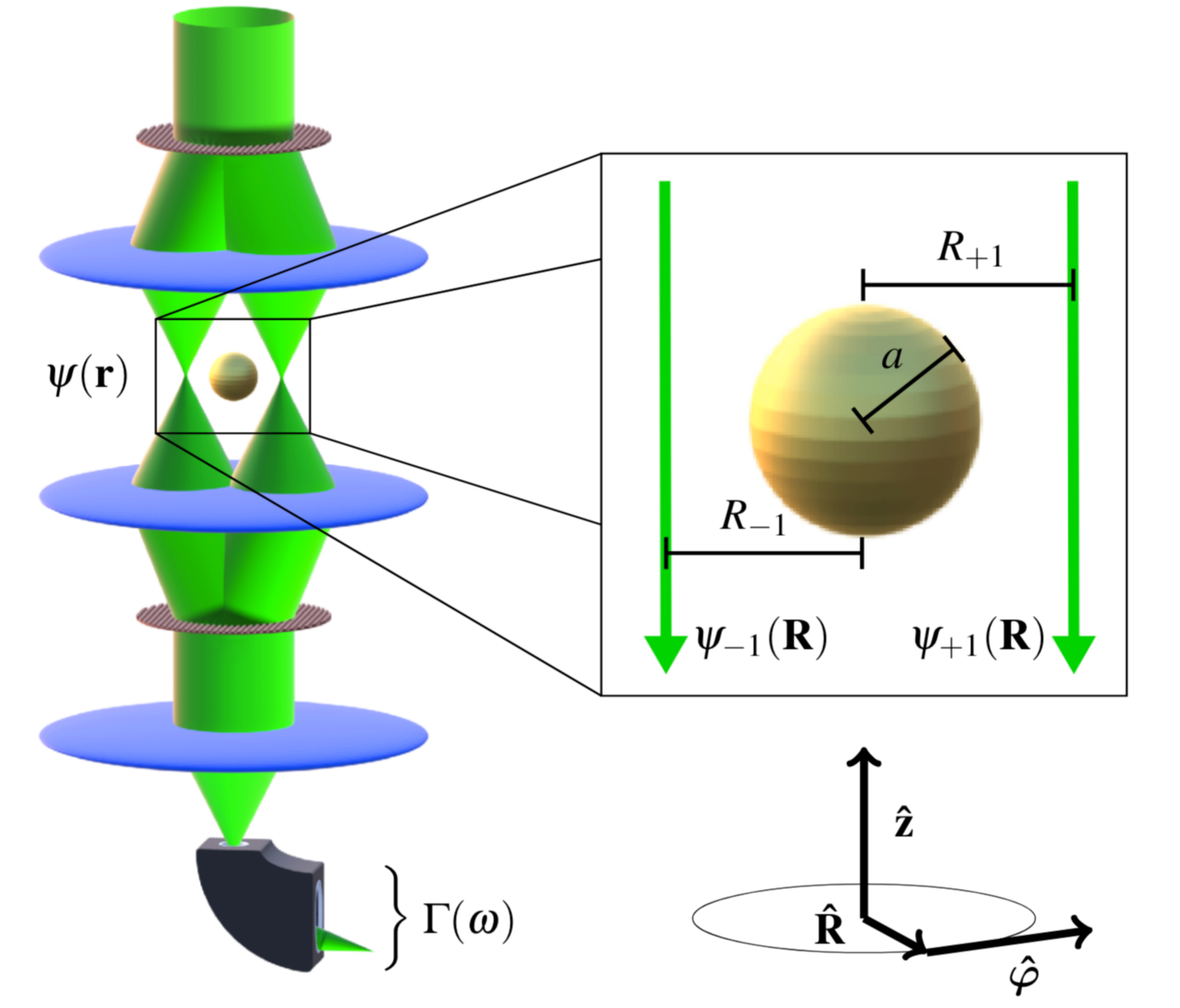}};
    \node[inner sep=0pt] at (7,0)
        {\includegraphics[height=2.5in]{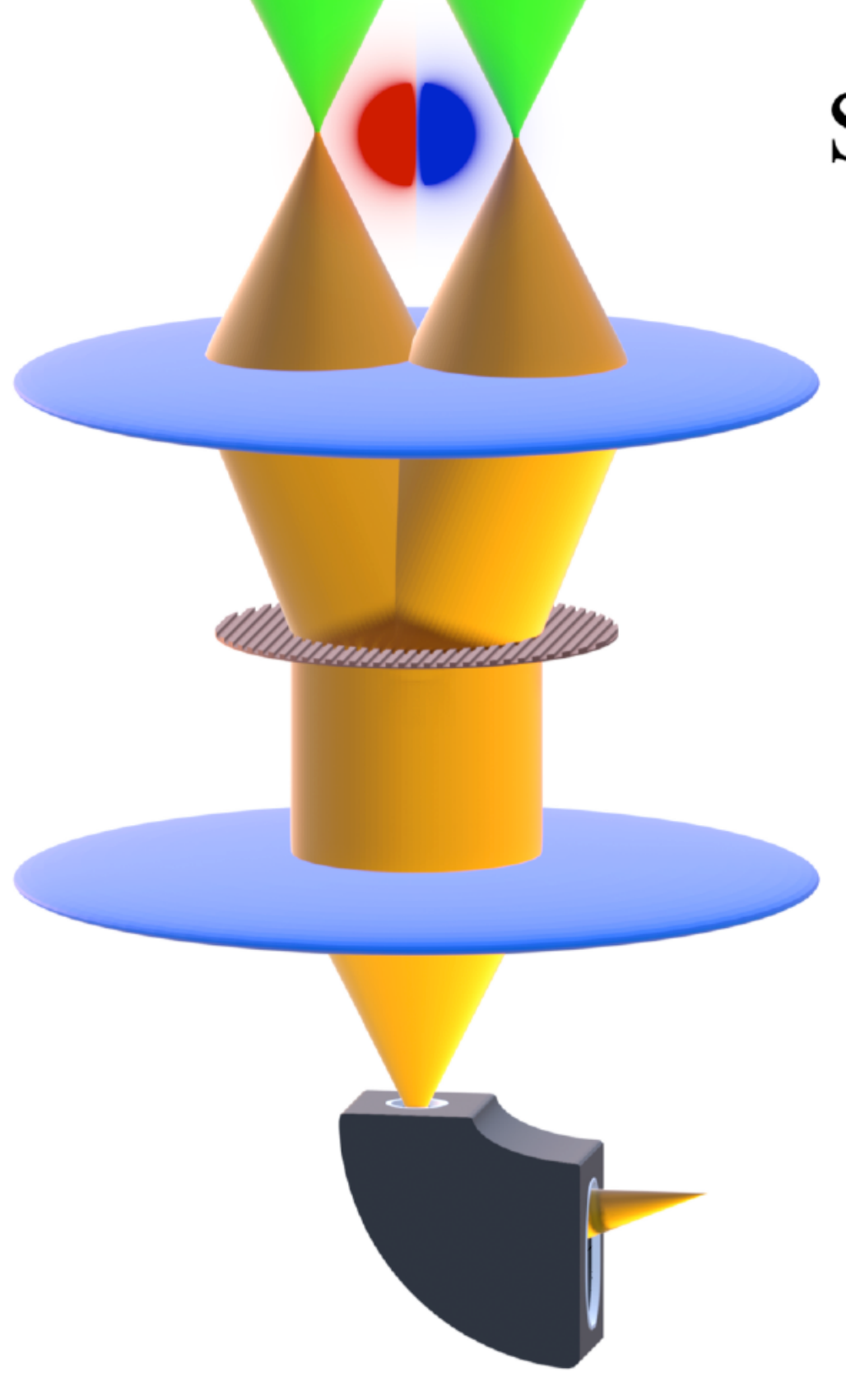}};
    \fill[white] (8.0,2.25) rectangle (9.55,3.75);
    \node[inner sep=0pt] at (8.75,2.55)
        {$\Gamma(\omega)$};
    \draw[-,dashed,line width=0.5]
        (7.4,2.55) -- (8.2,2.55);
    \node[inner sep=0pt] at (8.75,0.75)
        {$\frac{d\,\Gamma(\omega)}{d\mathbf{p}_\perp}$};
    \draw[-,dashed,line width=0.5]
    (7.85,0.3) -- (9.5,0.3);
    \node[inner sep=0pt] at (8.75,-0.2)
        {$\frac{d\,\Gamma_{\rm out}(\omega)}{d\mathbf{p}_\perp}$};
    \node[inner sep=0pt] at (9.1,-2.3)
        {$\Gamma_{\rm out}(\omega)|_0$};
    \draw[-,gray,line width=0.25]
        (6.4,0.2) -- (6.0,-0.825);   
    \draw[-,gray,line width=0.25]
        (7.55,0.2) -- (7.15,-0.825); 
    \draw[-,gray,line width=0.25]
        (6.0,-0.825) -- (6.575,-1.85);
    \draw[-,gray,line width=0.25]
        (7.15,-0.825) -- (6.575,-1.85);
    \draw[-,gray,line width=0.25]
        (6.4,0.2) -- (6.8,-0.825);   
    \draw[-,gray,line width=0.25]
        (7.55,0.2) -- (7.95,-0.825);
    \draw[-,gray,line width=0.25]
        (6.8,-0.825) -- (7.375,-1.85);
    \draw[-,gray,line width=0.25]
        (7.95,-0.825) -- (7.375,-1.85);
    \draw[-,gray,line width=0.25]
        (6.4,0.2) -- (7.2,-0.825);   
    \draw[-,gray,line width=0.25]
        (7.55,0.2) -- (8.35,-0.825);
    \draw[-,gray,line width=0.25]
        (7.2,-0.825) -- (7.775,-1.85);
    \draw[-,gray,line width=0.25]
        (8.35,-0.825) -- (7.775,-1.85);
    \draw[-,gray,line width=0.25]
        (6.4,0.2) -- (5.6,-0.825);   
    \draw[-,gray,line width=0.25]
        (7.55,0.2) -- (6.75,-0.825); 
    \draw[-,gray,line width=0.25]
        (5.6,-0.825) -- (6.175,-1.85);
    \draw[-,gray,line width=0.25]
        (6.75,-0.825) -- (6.175,-1.85);
    \node[scale=0.75,inner sep=0pt] at (5.3,-1.65)
        {$m=\cdots,$};
    \node[scale=0.75,inner sep=0pt] at (6.1,-1.65)
        {$-2,$};
    \node[scale=0.75,inner sep=0pt] at (6.5,-1.65)
        {$-1,$};
    \node[scale=0.75,inner sep=0pt] at (7.0,-1.65)
        {$0,$};
    \node[scale=0.75,inner sep=0pt] at (7.4,-1.65)
        {$1,$};
    \node[scale=0.75,inner sep=0pt] at (7.8,-1.65)
        {$2,$};
    \node[scale=0.75,inner sep=0pt] at (8.2,-1.65)
        {$\cdots$};
    \fill[white] (-1.6,-2.25) rectangle (-0.1,-3.75);
    \node[inner sep=0pt] at (-3.6,-3.1)
        {EELS};
    \node[inner sep=0pt] at (6,-2.5)
        {EELS};
    \node[inner sep=0pt] at (-3.8,3.5)
        {(a)};
    \node[inner sep=0pt] at (5.5,3.5)
        {(b)};
\end{tikzpicture}
\caption{\label{fig:fig3} (a) Geometry of two STEM probes, with wave function $\psi(\br) = (\psi_{-1}(\bR)+\psi_{+1}(\bR))\ee^{ik_zz}$, incident on a metallic NP in the 2GeMZI. (b) Diagram of inelastically scattered electrons propagating through the 2GeMZI. Energy-loss probability at the specimen, $\Gamma(\omega)$, transverse momentum-resolved loss probability immediately before and after the interferometer output grating, $\frac{d\Gamma(\omega)}{d\bp_{f\perp}}$ and $\frac{d\Gamma_{\rm out}(\omega)}{d\bp_{f\perp}}$ respectively, and the loss probability in the $m=0$ output order of the interferometer, $\Gamma_{\rm out}(\omega)|_0$. }
\end{figure}
\end{center}
\FloatBarrier

\section{Multiple Free-Electron Probes in a 2GeMZI}

The input grating of a 2GeMZI creates a superposition of vertical electron probes spatially separated in the transverse plane, each with a well-defined relative phase. The electron wave function thus admits a separation into an overall factor standing for a plane wave along the longitudinal direction $z$ times a superposition of foci along the transverse coordinates ${\bf R}=(x,y)$. More explicitly, $\psi(\br) = \ee^{ik_zz}\sum_jc_j\psi_j(\bR)$, where each transverse probe wave function can be approximated as a sharply peaked, normalized, and nonoverlapping Gaussian amplitude function $\psi_j(\bR)\approx \sqrt{2/\pi\Delta^2}\,\ee^{-|\bR-\bR_j|^2/\Delta^2}$ weighted in $\psi(\br)$ by coefficients $c_j = |c_j|\ee^{i\phi_j}$ that are proportional to the Fourier coefficients attributed to the periodic shape of the 2GeMZI input grating (see below). These coefficients satisfy the normalization condition $\sum_j|c_j|^2=1$ if $|\bR_j-\bR_{j'}|\gg\Delta$ for every $j\neq j'$.

It is illustrative to consider first the EELS probability for a multi-focus configuration in which no grating is placed after the specimen. If the inelastic signal is resolved in Fourier space (i.e., as a function of the electron wave vector ${\bf p}_{f\perp}$ after interaction), following a well-established quantum-mechanical formalism \cite{garcia_de_abajo_optical_2010}, we find
\begin{equation}
\label{Gammap0}
    \begin{split}
        \frac{d\Gamma(\omega)}{d\bp_{f\perp}} = & \,-\frac{e^2}{\pi^2\hbar}\mbox{Im}\left\{\int d^3\br\int d^3\br'\,\psi^*_\perp(\bR)\psi_\perp(\bR')\ee^{i\bp_{f\perp}\cdot(\bR-\bR')}\ee^{-i\omega(z-z')/v}\,G_{zz}(\omega,\bR,z,\bR',z')\right\},
    \end{split}
\end{equation}
where $\psi_\perp(\bR) = \sum_jc_j\psi_j(\bR)$ is the transverse wave function of the incident electron, whereas the transmitted electron wave function is selected as a transverse plane wave $\ee^{i\bp_{f\perp}\cdot\bR}$ of definite lateral momentum $\hbar\bp_{f\perp}$. The resulting momentum-resolved EELS probability reduces to
\begin{equation}
\label{Gammap}
    \begin{split}
        \frac{d\Gamma(\omega)}{d\bp_{f\perp}} = & \,-\frac{2e^2}{\pi^3\hbar\Delta^2}\sum_{j,j'}\mbox{Im}\bigg\{c_j^*c_{j'}\int d^2\bR\,\ee^{-|\bR-\bR_j|^2/\Delta^2}\ee^{i\bp_{f\perp}\cdot\bR}\int d^2\bR'\,\ee^{-|\bR'-\bR_{j'}|^2/\Delta^2}\ee^{-i\bp_{f\perp}\cdot\bR'} \\
        & \,\,\,\,\,\,\,\,\,\,\,\,\,\,\,\,\,\,\,\,\,\,\,\,\,\,\,\,\, \times \int dz \int dz'\, \ee^{-i\omega(z-z')/v}\,G_{zz}(\omega,\bR,z,\bR',z')\bigg\}.
    \end{split}
\end{equation}
This expression carries interference information from the different electron foci, which allows us to investigate the nonlocal optical response of the specimen, imprinted on the Green tensor for different locations $\bR\neq\bR'$. However, gaining access to such information requires measuring a narrow range of transverse momenta, as demonstrated in a recent experiment \cite{GBL17} that relied on post-selection over a small angular range of transmitted electrons. Indeed, when integrating Eq.\ (\ref{Gammap}) over $\bp_{f\perp}$, the interference is quickly washed out, with the factor $\ee^{i\bp_{f\perp}\cdot(\bR-\bR')}$ yielding a delta function $\delta(\bR-\bR')$, so that only information for $\bR=\bR'$ is retained and the result reduces to an incoherent sum of the EELS probability for the single-probe configuration, weighted by the electron probability distribution in the final transverse impact-parameter plane (i.e., a sum of the probability for independent probes). We remark that momentum-resolved interferometry based on Eq. (\ref{Gammap}) requires a drastic reduction of the electron signal by many orders of magnitude.

In contrast, in the 2GeMZI, the second grating guarantees a post-selection of the measured electron that offers maximum overlap with the incident electron. As we argue in the main text, the final (measured) electron state adopts the same form as the incident electron, so the interference signal has a magnitude similar to a regular single-probe EELS signal, only limited by the ability of the two gratings to transmit electrons along the first-order diffraction beams. The EELS probability is then given by an expression analogous to Eq.\ (\ref{Gammap0}), but with the transverse wave function $\ee^{i\bp_{f\perp}\cdot\bR}$ substituted by the final multi-focus wave function $\tilde\psi_\perp(\bR) = \sum_j\tilde c_j\psi_j(\bR)$, with coefficients $\tilde c_j$ determined by the second grating. More precisely,
\begin{equation}
    \begin{split}
        \Gamma(\omega)= & \,-\frac{e^2}{\pi^2\hbar}\mbox{Im}\left\{\int d^3\br\int d^3\br'\,\psi^*_\perp(\bR)\psi_\perp(\bR')\tilde\psi_\perp(\bR)\tilde\psi_\perp^*(\bR')\ee^{-i\omega(z-z')/v}\,G_{zz}(\omega,\bR,z,\bR',z')\right\},
    \end{split}
\end{equation}
which, for nonoverlapping focal spots, readily becomes
\begin{equation}
\label{GccccA}
    \begin{split}
        \Gamma(\omega) \approx & \sum_{j,j'}\mbox{Im}\left\{c_j^*c_{j'}\tilde c_j\tilde c_{j'}^*\, A(\omega,\bR_j,\bR_{j'})\right\},
    \end{split}
\end{equation}
where $A$ is the complex loss function defined in Eq.\ (\ref{AAA}).

Regarding the wave function coefficients $c_j$, in the weak-phase approximation, upon transmission through the first grating, the passing electron wave function $\psi$ is multiplied by a transmission phase-factor function $\ee^{i\eta t(\bR)}$, where $\eta$ denotes the phase shift and amplitude loss per unit thickness of the grating material, and $t(\bR)$ is the periodic thickness profile of the grating, which is a function of the transverse (along the grating plane) coordinates $\bR$. Since $t(\bR)$ is periodic, the whole transmission function can be written as the Fourier series $\ee^{i\eta t(\bR)} = \sum_j{c}_j\ee^{-i\bR\cdot\bp_{\perp,j}}$, where the sum runs over the reciprocal lattice vectors of the grating, $\bp_{\perp,j}$, each of them corresponding to a different focal spot of lateral position $\bR_j$ at the sample plane after focusing by the upper lens in the sketch of Fig.\ 2(b) in the main text. The coefficients $\tilde c_j$ produced by the second grating can be obtained by following a similar decomposition. In our setup, both gratings are designed with identical parameters, including the same pitch, so their associated coefficients are identical (up to fabrication and positioning imperfections), except for a lateral variable displacement ${\bf x}_0$ that we actively introduce in the second grating relative the first one, which yields $\tilde c_j=c_j\ee^{i{\bf x}_0\cdot\bp_{\perp,j}}$. This leads to the combined product $c_j^*c_{j'}\tilde c_j\tilde c_{j'}^*=|c_j^*c_{j'}\tilde c_j\tilde c_{j'}^*|\ee^{i\Delta\phi_{j,j'}}$, where  $\Delta\phi_{j,j'}={\bf x}_0\cdot(\bp_{\perp,j}-\bp_{\perp,j'})$. (Incidentally, this intrinsic phase difference must be supplemented by the contribution of static potentials, see below.) Inserting this result in Eq.\ (\ref{GccccA}), we find that the two-probe loss probability in the 2GeMZI then reduces to
\begin{equation}
    \begin{split}
        \Gamma_{\text{out}}(\omega)\big|_0 \propto &= \sum_{j}|c_j\tilde{c}_j|^2\Gamma_{\{\text{1p}\}}(\omega,R_j)+\sum_{j,j'}|c_jc_{j'}\tilde{c}_j\tilde{c}_{j'}|\Gamma_{\text{int}}(\omega,\bR_j,\bR_{j'},\Delta\phi_{j,j'}),
\end{split}
\end{equation}
where $\Gamma_{\{\text{1p}\}}(\omega,R)=\mbox{Im}\left\{A(\omega,\bR,\bR)\right\}$ is the single-probe EELS probability and $\Gamma_{\text{int}}(\omega,\bR,\bR',\Delta\phi)= \mbox{Im}\left\{\ee^{i\Delta\phi} A(\omega,\bR,\bR')\right\}$ is the contribution of interference between two different probes passing by $\bR$ and $\bR'$ with a relative grating-controlled relative phase $\Delta\phi$. For the sphere, combining this result with Eq.\ (\ref{GKCA}), we find the explicit expression
\begin{equation}
    \begin{split}
        \Gamma_{\rm int}(\omega,\bR_j,\bR_{j'},\Delta\phi_{j,j'}) = & \,\, \frac{e^2}{\hbar\omega c}\sum_{l=1}^\infty\sum_{m=-l}^lK_m\left(q_\gamma R_j\right)K_m\left(q_\gamma R_{j'}\right) C^E_{lm}\mbox{Im}\{t^E_l(\omega)\} \, \cos\left(\Delta\phi_{j,j'}+m\Delta\varphi_{j,j'}\right) \\ 
        = & \,\, \sum_{l=1}^\infty\Gamma_{l,{\rm int}}(\omega,\bR_j,\bR_{j'},\Delta\phi_{j,j'}),
    \end{split}
\end{equation}
where the last line implicitly defines a decomposition in multipolar contributions.

If the input and output gratings of the 2GeMZI are optimized to create only two probes (see Fig.\ \ref{fig:fig3}), we have the condition $c_{|m|\neq1}\approx\tilde{c}_{|m|\neq1}\approx0$, which leads to
\begin{equation}
    \begin{split}
        \Gamma_{\text{out}}(\omega)\big|_0 \propto & \,\, \Gamma_{\{\text{1p}\}}(\omega,R_{-1})+\Gamma_{\{\text{1p}\}}(\omega,R_{+1}) +2\Gamma_{\text{int}}(\omega,\bR_{-1},\bR_{+1},\Delta\phi),
    \end{split}
\end{equation}
where the subscripts $\pm$ refer to $j=\pm1$ and $\Delta\phi\equiv\Delta\phi_{-1,1}$.

\section{Spatial Inelastic Interference in the {2GeMZI} by Multipole Components}

\begin{center}
\begin{figure}[h]
\begin{tikzpicture}
    \node[inner sep=0pt] at (0,0)
        {\includegraphics[width=6in]{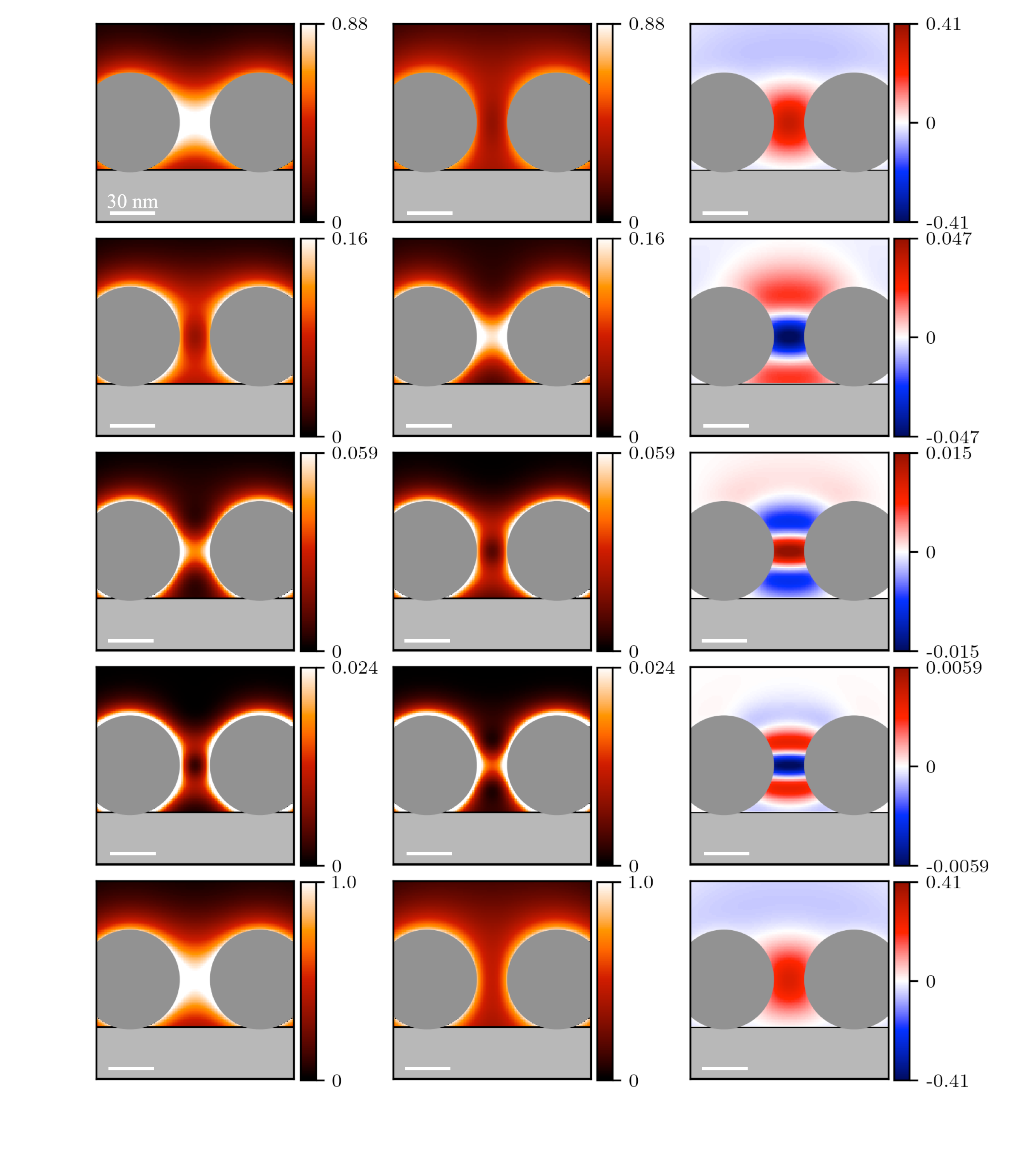}};
    \node[inner sep=0pt] at (-6.5,8.35)
        {(a)};
    \node[white,inner sep=0pt] at (-5.85,8.25)
        {(i)};
    \node[white,inner sep=0pt] at (-1.35,8.25)
        {(ii)};
    \node[inner sep=0pt] at (3.2,8.25)
        {(iii)};
    \node[inner sep=0pt] at (-6.5,5.1)
        {(b)};
    \node[white,inner sep=0pt] at (-5.85,5.0)
        {(i)};
    \node[white,inner sep=0pt] at (-1.35,5.0)
        {(ii)};
    \node[inner sep=0pt] at (3.2,5.0)
        {(iii)};
    \node[inner sep=0pt] at (-6.5,1.85)
        {(c)};
    \node[white,inner sep=0pt] at (-5.85,1.75)
        {(i)};
    \node[white,inner sep=0pt] at (-1.35,1.75)
        {(ii)};
    \node[inner sep=0pt] at (3.2,1.75)
        {(iii)};
    \node[inner sep=0pt] at (-6.5,-1.4)
        {(d)};
    \node[white,inner sep=0pt] at (-5.85,-1.5)
        {(i)};
    \node[white,inner sep=0pt] at (-1.35,-1.5)
        {(ii)};
    \node[inner sep=0pt] at (3.2,-1.5)
        {(iii)};
    \node[inner sep=0pt] at (-6.5,-4.6)
        {(e)};
    \node[white,inner sep=0pt] at (-5.85,-4.7)
        {(i)};
    \node[white,inner sep=0pt] at (-1.35,-4.7)
        {(ii)};
    \node[inner sep=0pt] at (3.2,-4.70)
        {(iii)};
\end{tikzpicture}
\caption{\label{fig:fig4} Simulated interferometric images of a single $a=30$\,nm gold NP over a 120$\times$120\,nm$^2$ scan region with a probe separation of 80\,nm. Spectra are integrated over the energy range 1-3 eV. Panels in rows (a-e) correspond to partial contributions of specific orbital angular momentum transfers: (a) $l=1$, (b) $l=2$, (c) $l=3$, (d) $l=4$, and (e) sum over $l=1-4$. Panels in different columns correspond to (i) $\Gamma_l(\omega)$ spectral images for $\Delta\phi_\text{int}=0$, (ii) $\Gamma_l(\omega)$ spectral images for $\Delta\phi_\text{int}=\pi$, and (iii) $\Gamma_{l,{\rm int}}(\omega)$ spectral images. All colorbar scales are normalized to the maximum intensity of (e)(i). The direction between the two-probe separation vector is parallel to the carbon support.}
\end{figure}
\end{center}

The relative phase difference between the two 2GeMZI probes receives contributions from both the interferometer alignment (the $\Delta\phi_{\rm int}\equiv\Delta\phi_{-1,1}={\bf x}_0\cdot(\bp_{\perp,-1}-\bp_{\perp,1})$ terms discussed above) and the external static potential traversed by the electron $\Delta\phi_{\rm ext}$, (see below):
\begin{equation}
    \Delta\phi = \Delta\phi_\text{int}+\Delta\phi_\text{ext}.
\end{equation}
Taking two interferometric images by scanning the probes over the NP with two interferometer alignments corresponding to $\Delta\phi_\text{int} = 0$ and $\Delta\phi_\text{int} = \pi$, we find
\begin{equation}
    \begin{split}
        \Gamma_{\rm out}(\omega,\Delta\phi_\text{int}=0)\big|_0 = & \,\,  \Gamma_{\{1\text{p}\}}(\omega,R_{-1})+\Gamma_{\{1\text{p}\}}(\omega,R_{+1})+2\Gamma_{{\rm int}}(\omega,\bR_{-1},\bR_{+1},\Delta\phi_\text{ext}), \\ 
        \Gamma_{\rm out}(\omega,\Delta\phi_\text{int}=\pi)\big|_0 = & \,\,  \Gamma_{\{1\text{p}\}}(\omega,R_{-1})+\Gamma_{\{1\text{p}\}}(\omega,R_{+1})-2\Gamma_{{\rm int}}(\omega,\bR_{-1},\bR_{+1},\Delta\phi_\text{ext}), \\ 
    \end{split}
\end{equation}
so the the interference component can be expressed as
\begin{equation}
    \begin{split}
        \Gamma_{{\rm int}}(\omega,\bR_{-1},\bR_{+1},\Delta\phi_\text{ext}) = \frac{1}{4}\left[\Gamma_{\rm out}(\omega,\Delta\phi_\text{int}=0)\big|_0-\Gamma_{\rm out}(\omega,\Delta\phi_\text{int}=\pi)\big|_0\right].
    \end{split}
\end{equation}
Given that the electron probes are composed of charged particles and can induce and deposit charge on the NP throughout a scan and that a modest amount of charge on the surface of the NP can cause a static potential resulting in a spatially dependent external phase shift between the probes, $\Delta\phi_\text{ext}$, we simulate interferometric images of the gold NP with a potential $V_z=v_{\rm NP}/R+v_{\rm edge}/(y-y_0)$ to approximate the projected potential phase shift applied to each probe for the $l=1-4$ multipole modes, originating in the surface charge on the NP and amorphous carbon support. These simulations include interferometer phase shifts $\Delta\phi_\text{int}=0$ and $\Delta\phi_\text{int}=\pi$ allowing us to also show $\Gamma_{l,{\rm int}}(\omega,\bR_{-1},\bR_{+1},\Delta\phi_\text{ext})$ with a potential assumed to have $v_{\rm edge}/v_{\rm NP}=30$ (see Fig.\ \ref{fig:fig4}). As expected, the $l=1$ dipole mode dominates, but there can be considerable contributions from higher-order modes, especially near the edge of the NP. 

\begin{center}
\begin{figure}[h]
\begin{tikzpicture}
    \node[inner sep=0pt] at (0,0)
        {\includegraphics[width=6in]{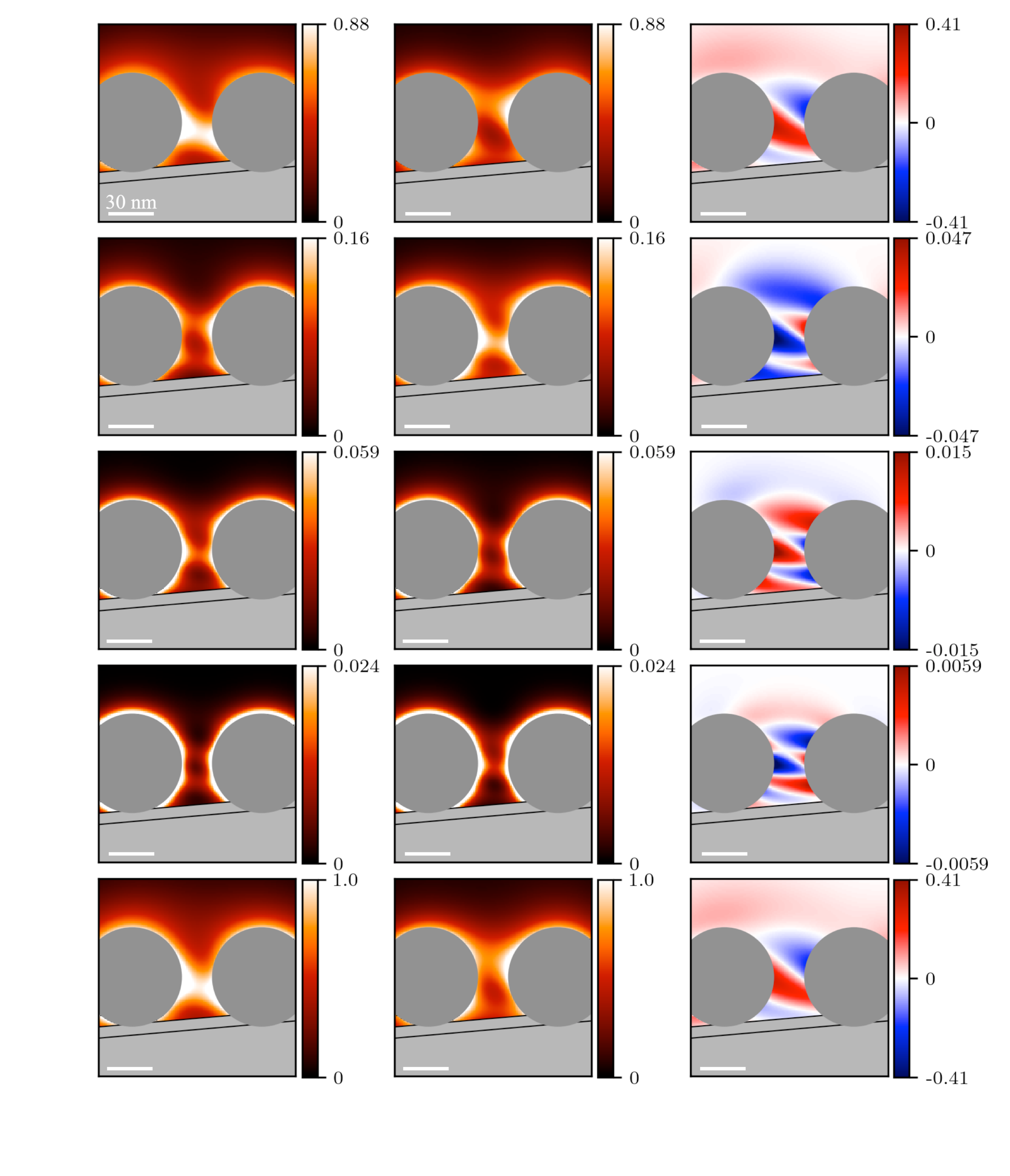}};
    \node[inner sep=0pt] at (-6.5,8.35)
        {(a)};
    \node[white,inner sep=0pt] at (-5.85,8.25)
        {(i)};
    \node[white,inner sep=0pt] at (-1.35,8.25)
        {(ii)};
    \node[inner sep=0pt] at (3.2,8.25)
        {(iii)};
    \node[inner sep=0pt] at (-6.5,5.1)
        {(b)};
    \node[white,inner sep=0pt] at (-5.85,5.0)
        {(i)};
    \node[white,inner sep=0pt] at (-1.35,5.0)
        {(ii)};
    \node[inner sep=0pt] at (3.2,5.0)
        {(iii)};
    \node[inner sep=0pt] at (-6.5,1.85)
        {(c)};
    \node[white,inner sep=0pt] at (-5.85,1.75)
        {(i)};
    \node[white,inner sep=0pt] at (-1.35,1.75)
        {(ii)};
    \node[inner sep=0pt] at (3.2,1.75)
        {(iii)};
    \node[inner sep=0pt] at (-6.5,-1.4)
        {(d)};
    \node[white,inner sep=0pt] at (-5.85,-1.5)
        {(i)};
    \node[white,inner sep=0pt] at (-1.35,-1.5)
        {(ii)};
    \node[inner sep=0pt] at (3.2,-1.5)
        {(iii)};
    \node[inner sep=0pt] at (-6.5,-4.6)
        {(e)};
    \node[white,inner sep=0pt] at (-5.85,-4.7)
        {(i)};
    \node[white,inner sep=0pt] at (-1.35,-4.7)
        {(ii)};
    \node[inner sep=0pt] at (3.2,-4.70)
        {(iii)};
\end{tikzpicture}
\caption{\label{fig:fig5} Same as Fig.\ \ref{fig:fig4}, but with the direction between the two-probe separation vector tilted $5^\circ$ relative to the carbon support.}
\end{figure}
\end{center}

When the probe separation direction is parallel to the carbon support, the $1/y$ part of the projected potential exactly cancels in the phase difference due to the mirror symmetry. Introducing an angular misalignment between the probe separation direction and the carbon support tilts and skews the interference fringes in a manner that is qualitatively consistent with the experiments. Figure\ \ref{fig:fig5} shows this effect for a 5$^\circ$ angular misalignment. 

Figures\ \ref{fig:fig4} and \ref{fig:fig5} confirm two key assumptions about the two-probe inelastic interference from a LPR in a single gold NP in the 2GeMZI when the probes are scanned on opposing sides of the NP in an aloof configuration. First, we observe that the dipolar optical response of the NP is dominant, so that the dipolar component of the inelastic signal is both qualitatively and quantitatively equivalent to the total signal. Second, an arbitrary external potential distorts the spatial interference fringes, but such fringes are still $2\pi$ periodic in the relative probe phase and conversely related to the elastic interference out of the interferometer over the scan region.

\FloatBarrier

\section{Inelastic Interferometry on a Second Gold Nanoparticle} 

\begin{center}
\begin{figure}[h]
\begin{tikzpicture}
    \node[inner sep=0pt] at (0,0)
        {\includegraphics[width=6in]{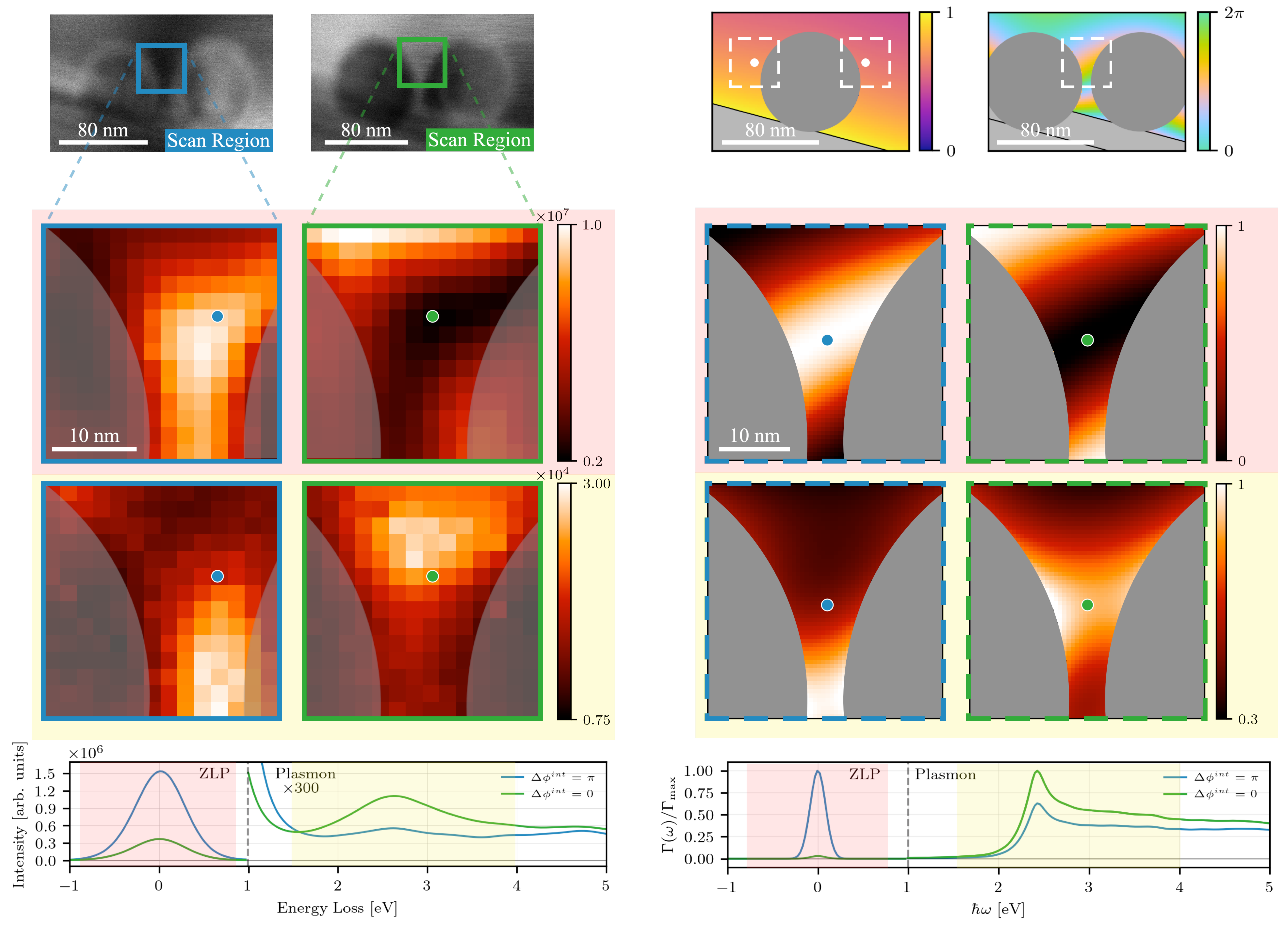}};
    \node[white,inner sep=0pt] at (-6.75,5.1) {(a)};
    \node[white,inner sep=0pt] at (-3.7,5.1) {(b)};
    \node[white,inner sep=0pt] at (-6.8,2.5) {(c)};
    \node[inner sep=0pt] at (-6.5,-3.8) {(d)};
    \node[white,inner sep=0pt] at (1.05,5.15) {(e)};
    \node[white,inner sep=0pt] at (4.35,5.15) {(f)};
    \node[white,inner sep=0pt] at (1.0,2.5) {(g)};
    \node[inner sep=0pt] at (1.4,-3.8) {(h)};
    \node[inner sep=0pt] at (-4.2,3.425) {Experiment};
    \node[inner sep=0pt] at (-5.7,3.15) {$\Delta\phi_\text{int}=\pi$};
    \node[inner sep=0pt] at (-2.7,3.15) {$\Delta\phi_\text{int}=0$};
    \node[inner sep=0pt] at (3.7,3.425) {Theory};
    \node[inner sep=0pt] at (2.2,3.175) {$\Delta\phi_\text{int}=\pi$};
    \node[inner sep=0pt] at (5.3,3.175) {$\Delta\phi_\text{int}=0$};
    \node[rotate=90,inner sep=0pt] at (-0.55,1.5) {ZLP Integrated};
    \node[rotate=90,inner sep=0pt] at (-0.55+7.75,1.5) {ZLP Integrated};
    \node[rotate=90,inner sep=0pt] at (-0.55,1.5-3.15) {Plasmon Integrated};
    \node[rotate=90,inner sep=0pt] at (-0.55+7.75,1.5-3.15) {Plasmon Integrated};
    
\end{tikzpicture}
\caption{\label{fig:fig6} (a,b) 2GeMZI bright field images
for a gold NP of 35\,nm radius under destructive (a) and constructive (b) interferometer alignments, showing
spectral image scan regions. (c) Experimental spectral images. (d) EELS spectra corresponding to the blue and green dotted scan locations in (c). (e) Simulated static projected potential $V_z(x,y)/V_{z,max}$. (f) Two-probe relative external phase due to the simulated projected potential, $\Delta\phi_\text{ext}$, for the metallic NP on a carbon support. (g) Simulated spectral images in the dashed box scan region of (f) for destructive and constructive interferometer alignments. (h) Simulated spectra corresponding to the blue and green dotted scan locations in (g).}
\end{figure}
\end{center}

Additional experimental data on a second gold NP of 35\,nm radius is offered in Fig.\ \ref{fig:fig6} here for a 80\,nm probe separation. The angular misalignment between the probe diffraction direction and the carbon support is larger and in an opposite direction compared to the dataset presented in the main text. This causes the tilt and skew in the projected potential phase difference to also be in the opposite direction, consistent with charging from the carbon support. This data also displays the converse interference relation between the ZLP and plasmon signals and is qualitatively consistent with the assumed form of the external potential and how the orientation of the scanned probes distorts the spatial interference fringes. 

\FloatBarrier

\bibliographystyle{apsrev4-2}
\bibliography{ms.bib}